\documentclass{aastex631}

\usepackage{rotating}
\received{2021 April 1}
\revised{2021 May 12}
\accepted{2021 May 28}
\shorttitle{The $\rm ^{12}C+^{12}C$ nuclear cross section and the compactness}
\shortauthors{Chieffi et al.}
\graphicspath{{./}{figures/}}
\newcommand{\msun}{$\rm M_\odot$}

\newcommand\nuk[2]{$\rm ^{\rm #2} #1$}

\begin{document}

\title{The impact of the new measurement of the $\rm ^{12}C+^{12}C$ fusion cross section on the final compactness of the massive stars}

\author[0000-0002-3589-3203]{Alessandro Chieffi}
\affiliation{Istituto Nazionale di Astrofisica - Istituto di Astrofisica e Planetologia Spaziali, Via Fosso del Cavaliere 100, I-00133, Roma, Italy}
\affiliation{Monash Centre for Astrophysics (MoCA), School of Mathematical Sciences, Monash University, Victoria 3800, Australia}
\affiliation{INFN. Sezione di Perugia, via A. Pascoli s/n, 06125 Perugia, Italy}
\email{Alessandro Chieffi: alessandro.chieffi@inaf.it}

\author[0000-0003-0390-8770]{Lorenzo Roberti}
\affiliation{Dipartimento di Fisica, ``Sapienza'' Università di Roma, Piazzale Aldo Moro 5, 00185, Rome, Italy}
\affiliation{Istituto Nazionale di Astrofisica - Osservatorio Astronomico di Roma, Via Frascati 33, I-00040, Monteporzio Catone, Italy}
\email{Lorenzo Roberti: lorenzo.roberti@inaf.it}

\author[0000-0003-0636-7834]{Marco Limongi}
\affiliation {INFN. Sezione di Perugia, via A. Pascoli s/n, 06125 Perugia, Italy}
\affiliation{Istituto Nazionale di Astrofisica - Osservatorio Astronomico di Roma, Via Frascati 33, I-00040, Monteporzio Catone, Italy}
\affiliation{Kavli Institute for the Physics and Mathematics of the Universe, Todai Institutes for Advanced Study, the University of Tokyo, Kashiwa, Japan 277-8583 (Kavli IPMU, WPI)}       
\email{Marco Limongi: marco.limongi@inaf.it}

\author[0000-0002-1819-4814]{Marco La Cognata}
\affiliation {INFN, Laboratori Nazionali del Sud, Via S. Sofia 62, 95123 Catania, Italy}
\email{Marco La Cognata: lacognata@lns.infn.it}

\author[0000-0002-4055-0811]{Livio Lamia}
\affiliation {INFN, Laboratori Nazionali del Sud, Via S. Sofia 62, 95123 Catania, Italy}
\affiliation{Dipartimento di Fisica, Università degli Studi di Catania, Via S. Sofia 64, 95123, Catania, Italy}
\email{Livio Lamia: llamia@lns.infn.it}

\author[0000-0001-5386-8389]{Sara Palmerini}
\affiliation {INFN. Sezione di Perugia, via A. Pascoli s/n, 06125 Perugia, Italy}
\affiliation{Dipartimento di Fisica e Geologia, Università degli Studi di Perugia, via A. Pascoli s/n, 06125 Perugia, Italy}
\email{Sara Palmerini: sara.palmerini@pg.infn.it}

\author[0000-0003-2436-6640]{Rosario Gianluca Pizzone}
\affiliation {INFN, Laboratori Nazionali del Sud, Via S. Sofia 62, 95123 Catania, Italy}
\email{Rosario Gianluca Pizzone: rgpizzone@lns.infn.it}

\author[0000-0002-9986-1518]{Roberta Spartà}
\affiliation {INFN, Laboratori Nazionali del Sud, Via S. Sofia 62, 95123 Catania, Italy}
\email{Roberta Spartà: rsparta@lns.infn.it}

\author[0000-0002-6953-7725]{Aurora Tumino}
\affiliation {INFN, Laboratori Nazionali del Sud, Via S. Sofia 62, 95123 Catania, Italy}
\affiliation{Facoltà di Ingegneria e Architettura, Università degli Studi di Enna "Kore", Cittadella Universitaria, 94100, Enna, Italy}
\email{Aurora Tumino: tumino@lns.infn.it}

\correspondingauthor{Alessandro Chieffi}

\begin{abstract}
We discuss how the new measurement of the \nuk{C}{12}+\nuk{C}{12} fusion cross section carried out with the Trojan Horse Method \citep{thm18} affects the compactness of a star, i.e. basically the binding energy of the inner mantle, at the onset of the core collapse. In particular, we find that this new cross section significantly changes the dependence of the compactness on the initial mass with respect to previous findings obtained in \cite{cl20} by adopting the classical cross section provided by \cite{cf88}. A non monotonic but well defined behavior is confirmed also in this case and no scatter of the compactness around the main trend is found. Such an occurrence could impact the possible explodability of the stars. 
\end{abstract}

\section{Introduction} \label{sec:intro}
Gravity, the fundamental force that acts in a star, and the energy losses, the driver of its temporal evolution, are two partners hugging each other in a {\it valzer} that ends when a third wheel shows up, either an electron degeneracy strong enough to fully counterbalance the gravity, or the onset of the collapse of the inner core. In the first case a star ends up as a white dwarf while in the second case a fraction of the star is returned to the interstellar medium and the remaining part remains locked in a remnant (neutron star or black hole). The relative fraction between these two components (remnant and ejecta) depends in most cases on the competition between the capability of the star to exploit a sufficient fraction of the energy carried by the escaping neutrinos and the binding energy of the mantle, that obviously opposes to its ejection. For sake of completeness, let us remind that in some mass intervals a fourth wheel shows up: stars between 8-10 \msun~ (this mass interval must be considered very indicative) form strongly electron degenerate cores but reach densities high enough for electron captures on \nuk{Ne}{20} and \nuk{Mg}{24} \citep{mi80,no84} to activate and to trigger the explosion while stars having He core masses M(He) between 40 and 130\msun~ enter the region in which a copious production of pairs $\rm e^++e^-$ due to $\gamma$'s annihilation softens the equation of state leading therefore to their explosion (pulsational pair instability supernovae if M(He)$<$64 and pair instability supernovae for M(He)$>$64) \citep{hw02}.

Self consistent 3D simulations of the explosion of a massive star is currently carried out by several groups, e.g. \cite{ja17,mu17,bu19}. These simulations are extremely demanding in terms of computer power, even on massively parallel computers, and hence only a limited number of simulations have been performed so far. For this reason several groups \citep{oo11,oo13,er16,su18} tried to adopt a different strategy; the idea is to firstly compute the explosion of a large set of stellar models with a 1D hydro code, and then try to correlate the successfulness or failure of an explosion to one, or few, properties of the pre-supernova models. In this way it would be possible to predict the final fate of a star without the effort of following the explosion. Such a strategy was first proposed in \cite{oo11,oo13} who found a reasonable correlation between the explodability of a star and a single parameter that they named the $\xi$ parameter, $\xi_{\rm M_k}$=$\rm M_k$(\msun)$\rm / R_{@M_k}$(1000~km); they found that the best mass location where to evaluate this parameter is 2.5\msun. However, it was shown in \cite{er16} that a criterion based on a single parameter is not sufficient to determine the explodability of a star and it was proposed the use of two parameters: $\xi$  evaluated at 2.5\msun~ and $\mu_4$, a parameter correlated to the density gradient present at the base of the O shell.
The important role played by the structure of a star at the onset of the core collapse to get a successful explosion or not (sometimes not considered as fundamental as the study of the neutrino transport) lead \cite{su18} to recompute a wide set of stellar models from the main sequence up to the core collapse and to use the two parameter criterion proposed by \cite{er16} to identify the stars that successfully explode from those that fully collapse in a remnant. Among the various results, this paper showed a large scatter in the $\xi$ parameter even for variations of the initial mass as low as 0.01\msun.  

Quite recently we, \citep{cl20}, hereinafter CL20, reanalyzed in great detail the dependence of the compactness parameter $\xi$ on the initial mass without attempting an a priori determination of the possible explodability of the stars. The main reason was that we were mainly interested in understanding if the structure of a star at the onset of the core collapse is "robust" or it is sometimes the result of some random, fortuitous things that occur within the star. In addition to this, it is worth to remind that the calibrations proposed by both \cite{oo11} and \cite{er16} were based on explosions computed by adopting 1D hydro codes (calibrated on some observed Supernova, like e.g. SN1987A), and not with state of the art 3D auto-consistent explosions. In fact, \cite{bu19} warned against the adoption of simple parameters evaluated at the supernova stage to infer the explodability of a star.

In our previous work we showed there is no random change, at the pre-supernova stage, of the $\xi_{2.5}$ parameter of a star with the initial mass, rather a well defined non monotonic trend with the mass is present and well understood. In addition to this, we showed the crucial role played by C burning in sculpting the relation between the initial mass and the $\xi$ parameter, that synthetically quantifies the binding energy of the mantle of a star. 

The analysis performed by CL20 was obtained by adopting an efficiency of the \nuk{C}{12}+\nuk{C}{12} nuclear reaction rate based on the amount of \nuk{C}{12} left by the He burning (that is the result of the interplay among 3$\alpha$, \nuk{C}{12}($\alpha$,$\gamma$)\nuk{O}{16} and the instabilities, thermal and mechanical) and the nuclear cross section provided by \cite{cf88}. This nuclear cross section has never been considered well robustly established at energies of astrophysical interest, due to the complex structure of the \nuk{C}{12} nucleus. Sensitivity study aimed to explore which would have been the consequence of a variation of this nuclear cross section on the evolution of a massive star were presented in a few papers. \cite{ga07} presented a first exploratory analysis of the effects of a strong reduction of this nuclear cross section on the evolution of a few massive stars. Later on, \cite{be12} and \cite{pi13} explored the consequences of both an increase and a decrease of this nuclear cross section on the evolution of a set of massive stars (15, 20, 25, 32 and 60\msun). None of this paper was based on a new measurement of this nuclear reaction but either the cross section provided by \cite{cf88} was multiplied or divided by a fixed factor or a new cross section was derived by assuming the presence of a strong resonance at a center of mass energy of 1.5 MeV. Since a new measure of this pivotal nuclear cross section has been recently published by \cite{thm18} and since we showed very recently how important is the C burning for the final compactness (and hence explodability) of a star, we thought it was the right time to show how the results presented by CL20 are modified by the adoption of this new measurement of the \nuk{C}{12}+\nuk{C}{12} nuclear cross section.

The paper is organized as follows: Section 2 is devoted to the presentation of the new nuclear cross section and Section 3 to the FRANEC code. Section 4 discusses the central C burning phase while Section 5 addresses the shell C burning and its role in sculpting the final compactness of a star. A final discussion and conclusions follow.

\section{The new \nuk{C}{12}+\nuk{C}{12} nuclear cross section}\label{sec:ncs}

\begin{figure}[ht!]
\epsscale{1.}
\plotone{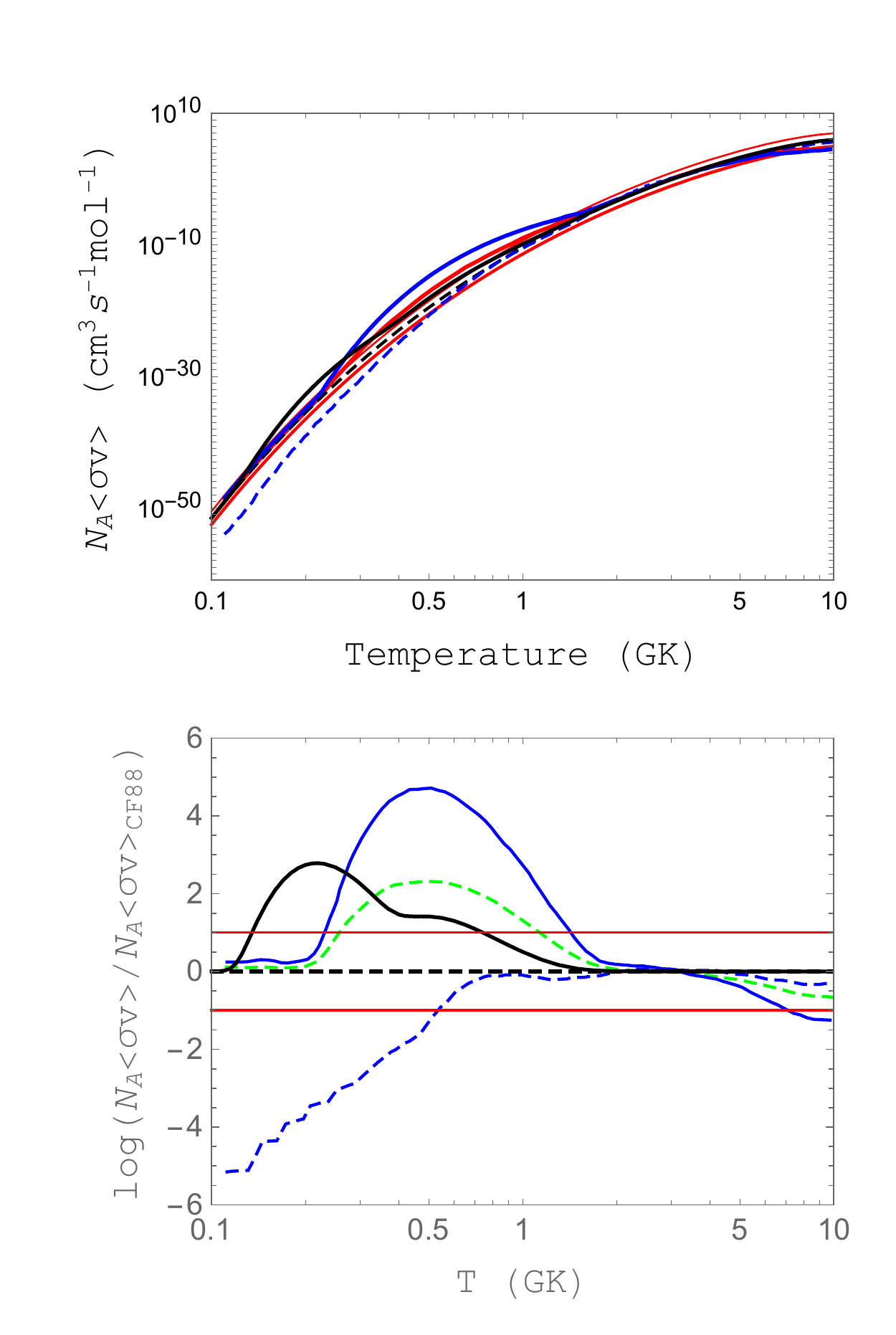}
\caption{Top panel: the THM \nuk{C}{12}+\nuk{C}{12} reaction rate from \citep{thm18} used in this study (black solid line) compared to the CU rate (blue solid line) from \citep{be12,pi13}, CI rate (green dashed line) from \citep{be12}, CL rate (blue dashed line) from \citep{pi13}, CF88 rate (black dashed line) from \citet{cf88}, CF88t10 and CF88d10 rates (red thinner and thicker lines) from \citep{pi13}
Bottom panel: the rates in the top panel are shown normalized to CF88. 
\label{fig:rateratio}}
\end{figure}

Astrophysical models often use the reaction rate of \citet{cf88} (CF88) as a reference in nucleosynthesis networks. 
It is based on the experimental data published no later than 1981, reaching down to a center of mass energy of
about 2.5~MeV, far from the 2~MeV upper edge of the astrophysical energy interval of interest.
Measuring below such energies is indeed an extremely difficult task, never achieved so far since at 2~MeV cross sections of the relevant p and $\alpha$ channels
are already as small as picobarns. This rate therefore relies on extrapolation after available 
experimental data from 2.5 to 6.5 MeV have been fitted to an astrophysical factor given by $S(E) = 3\times 10^{16} \exp(-0.46\,E)$~MeV-b. This is a very simple formula ignoring the effects of possible low-lying resonances.
As mentioned above, \citet{be12,pi13} have heuristically explored how even a single low-lying resonance can dramatically impact the evolution and nucleosynthesis of massive stars. In particular, \citet{be12} uses an ‘upper limit’ rate (CU) $\sim$50,000 times higher than CF88 at 0.5 GK and an intermediate rate (CI), representing the geometric mean of CF88 and CU. The calculations by \citet{pi13} are done with the CU rate and with a lower rate limit (CL) $\sim$20 times lower than CF88 at 0.5 GK. Variations of CF88 by a factor of 10 (CF88T10) and 0.1 (CF88d10) are also considered. Recent direct measurements \citep{ji18,fr20,ta20} and calculations \citep{bn20} refer that the astrophysical factor is not smooth, yet it shows a rich pattern of resonances. Also, traditional approaches such as the debated hindrance model seem to be excluded by more refined microscopic calculations \citep{go19}. However, p and $\alpha$ reaction channels with the recoil nuclei in their ground (p$_0$,$\alpha_0$) states are not accessible via direct measurements employing the particle-$\gamma$ coincidence. This technique was used in all recent direct experiments at centre-of-mass energies below 3 MeV to reduce the background contribution.
To overcome the experimental limits of direct measurements and to avoid the uncertain extrapolation, the Trojan Horse Method (THM) was recently applied to study the $^{12}$C$+^{12}$C fusion \citep{thm18} at centre-of-mass energies from 2.7 down to 0.8~MeV, including the whole energy interval of astrophysical interest and a region of overlap with direct data for normalization. 
$^{14}$N was used to transfer a $^{12}$C nucleus inside the nuclear field of a $^{12}$C target and the modified R-matrix approach was applied to determine the $^{12}$C$+^{12}$C astrophysical S(E) factor unhindered by Coulomb barrier effects. Thanks to the unique features of this indirect method  (see \citet{spit2019,tumino21} for a review), the measured astrophysical S(E) factor is also free of electron screening or background effects,
due, for instance, to contaminants in the target, which play a critical role in direct
experiments (see, e.g., \citet{fr20}). 
The THM has been successfully applied in the last decades to investigate crucial reactions also in other astrophysical contexts, such as primordial nucleosynthesis (\citet{lamia19,sparta20} and references therein), 
novae \citep{laco17} and AGB nucleosynthesis \citep{Pizzone17}. 
In the present case, the THM data shows clearly resolved low lying resonance structures in the astrophysical factor, 
later supported by direct
measurements in the overlapping energy region \citep{fr20} and by calculations \citep{bn20}. The THM measurement refers to both  
p and $\alpha$ reaction channels with the recoil nuclei in their ground (p$_0$,$\alpha_0$) and first excited states (p$_1$,$\alpha_1$),
which play a major role at astrophysical energies. Results from \citet{ta20} at centre-of-mass energies below 3 MeV provide a much lower upper limit for the p$_1$ and $\alpha_1$ channels, in significant disagreement with \citet{fr20}. In an overall comparison of recent results (THM and direct) in terms of total cross section \citep{tumino21}, agreement within experimental errors is observed in the whole energy region of astrophysical relevance, except for the two lowest data points of \citet{Spillane07} and those from \citet{ta20} at centre-of-mass energies $2.7 - 3\; {\rm MeV}$. The reaction rate was evaluated using standard formulas and fitted to an analytical function with an accuracy better than 0.7\% \citep{thm18}. Figure~\ref{fig:rateratio} shows in its top panel the \nuk{C}{12}+\nuk{C}{12} THM rate (black solid line) compared with CF88 (black dashed line) as well as with CU (blue solid line), CI (green dashed line), CL (blue dashed line) CF88T10 (red thinner solid line) and CF88d10 (red thicker solid line). In the bottom panel, all the rates of the top panel are shown normalized to CF88. None of the explorations are confirmed by the THM rate that experiences a smoother increase than CU and CI, crossing CF88T10 around 0.75 GK while reaching a much lower value than CU around 0.5 GK. Below 0.4 GK the THM rate shows a further increase by up to a factor of $\sim$800 owing to the lowest-energy resonances occurring at center of mass energies around 1 MeV. Within a comparison with CF88, in the temperature range of interest for the current calculations, the increasing factor of the THM rate varies between $\sim25$ at T=0.5~GK (Log\,T=8.7) and $\sim3$ at T=1~GK (Log\,T=9).

\section{The models}
The release of the FRANEC code adopted in the present computations, as well as all the input physics, is the same used in CL20. The only difference is the adoption of the new cross section for the \nuk{C}{12}+\nuk{C}{12} nuclear reaction (see Section \ref{sec:ncs}). The grid of models ranges between 12 and 27\msun~ with a step in mass of 0.1\msun. This grid is sufficient to show how the new nuclear reaction rate affects the advanced burning phases of the massive stars and hence their compactness at the onset of the core collapse. Obviously, since these models differ with respect to our previous ones only for the nuclear reaction rate of the \nuk{C}{12}+\nuk{C}{12}, their internal structure is identical to that of the models already published up to the end of the central He burning phase, so that both the amount of C left by the He burning and the initial size of the CO core mass are the same of the reference models. For simplicity, in the following the reference (i.e. those computed with the \nuk{C}{12}+\nuk{C}{12} nuclear reaction rate provided by CF88) and the new (i.e. those computed with the rate presented in Section \ref{sec:ncs}) models will be labelled CF88 and THM. 
For sake of clearness, let us remind the reader that throughout this paper we will adopt as proxy for the compactness, the $\xi$ parameter as originally defined by \cite{oo11}, i.e.: $\xi_{2.5}$=2.5\msun$\rm / R_{2.5}$(1000~km).

\section{The central carbon burning} \label{sec:carcen}
\begin{figure}[ht!]
\epsscale{1.1}
\plotone{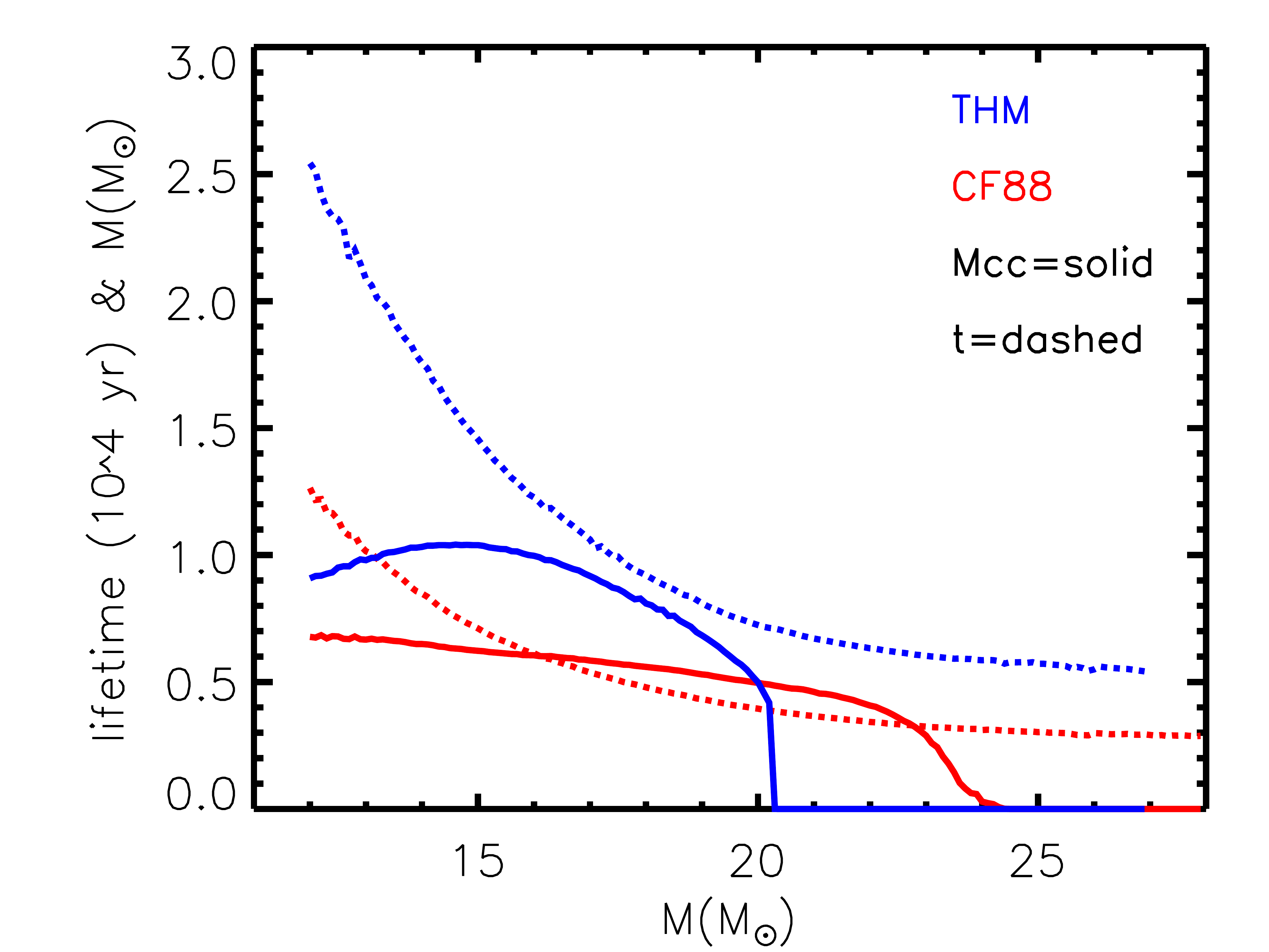}
\caption{Maximum size of the convective core (solid lines) and lifetime in central C burning (dashed lines) for the CF88 (red) and THM (blue) models. \label{fig:cbur}}
\end{figure}

Figure \ref{fig:cbur} shows a comparison between the CF88 models (red lines) and the THM ones (blue lines). The solid and dashed lines show, respectively, the maximum size of the convective core and the lifetime of the stars in central C burning, as a function of the initial mass. The adoption of the new nuclear reaction rate roughly doubles the lifetime of a star in central C burning, increases the maximum size of the convective core in stars less massive than, say, 20\msun, and lowers the maximum mass that forms a convective core down to about 20\msun. These results are the direct consequence of the fact that C burning occurs in an environment where strong neutrino energy losses occur. In fact, if the neutrino energy losses were not present, the increase of the nuclear reaction rate of the \nuk{C}{12}+\nuk{C}{12} would have been simply counterbalanced by a temperature drop, keeping the energy production rate at the level required by the CO core mass size. This can be easily seen by computing the evolution of a 15\msun~ without the inclusion of the neutrino energy losses. In this case the maximum size of the convective core in C burning varies from 1.845\msun~ (CF88) to 1.823\msun~ (THM) while the lifetime of the central C burning changes from $1.14\times10^{5}$ yr to $1.08\times10^{5}$ yr. Extremely small differences.

\begin{figure}[ht!]
\epsscale{1.1}
\plotone{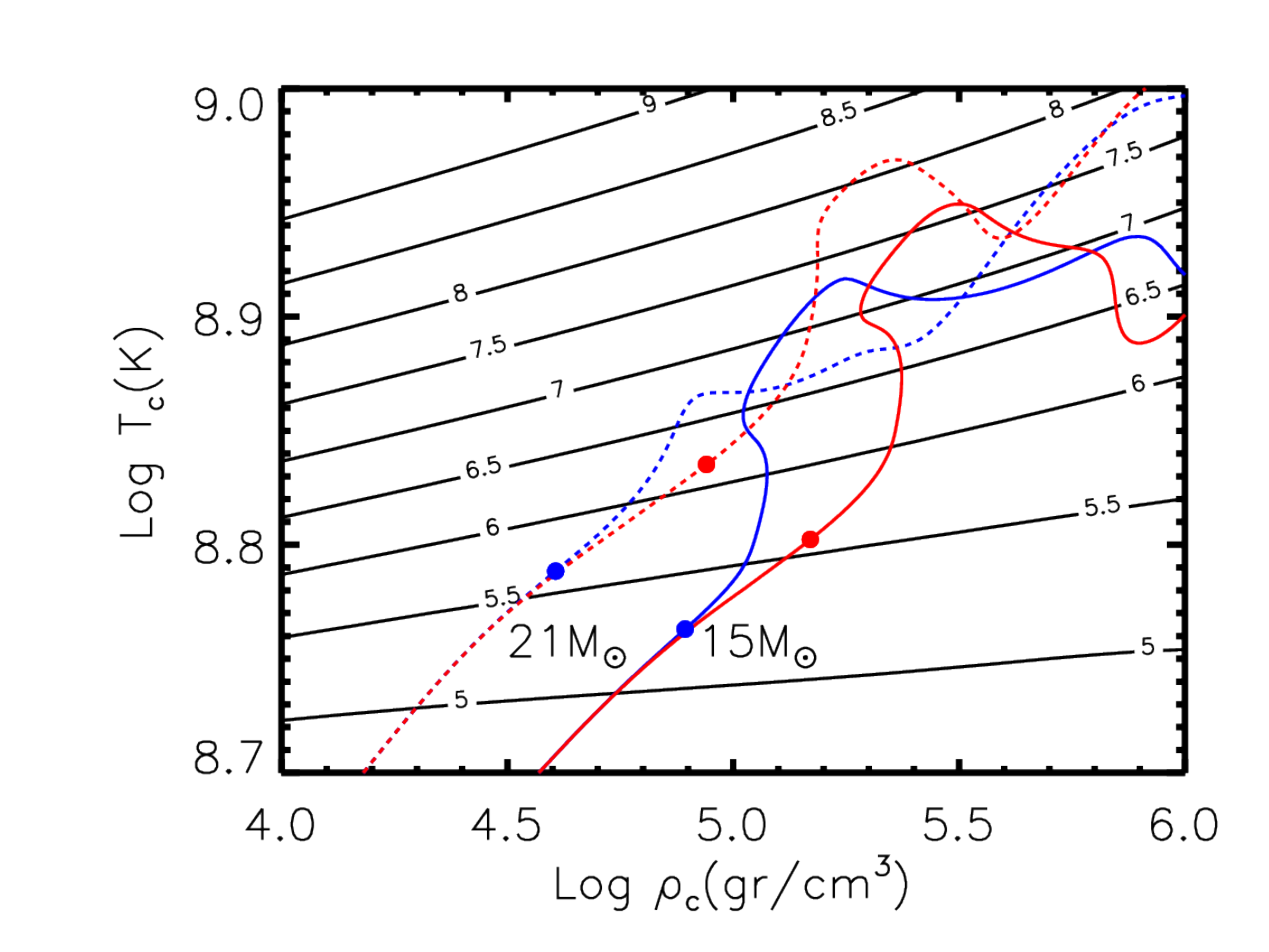}
\caption{Logarithm of the iso neutrino energy losses ($\rm erg~g^{-1}s^{-1}$) in black. Path followed by the 15\msun~ (solid lines) and the 21\msun~ (dashed lines). The THM and the CF88 models are shown in blue and red, respectively. \label{fig:conftcroc}}
\end{figure}

\begin{figure}[ht!]
\gridline{\fig{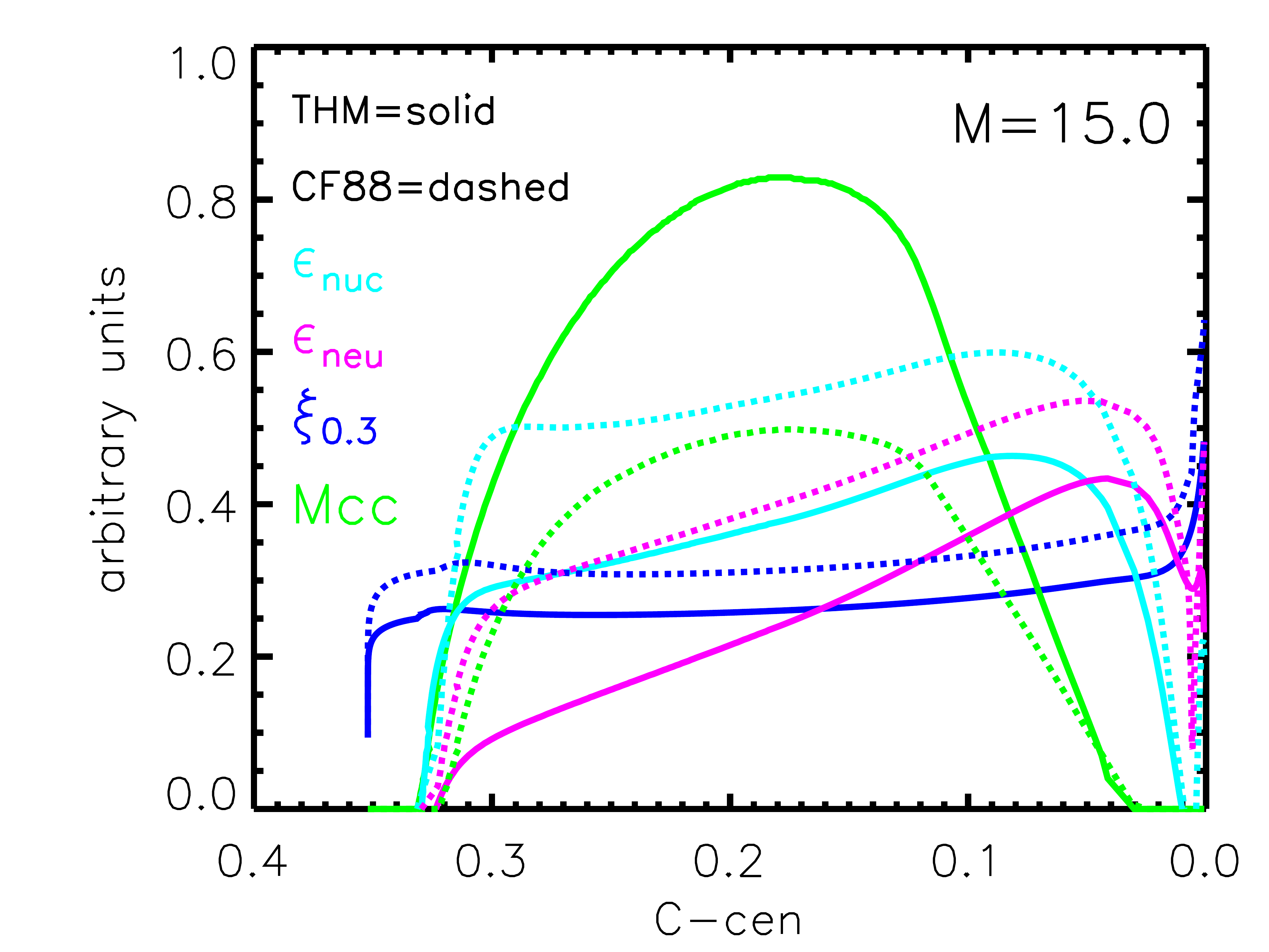}  {0.45\textwidth}{(a)}}
\gridline{\fig{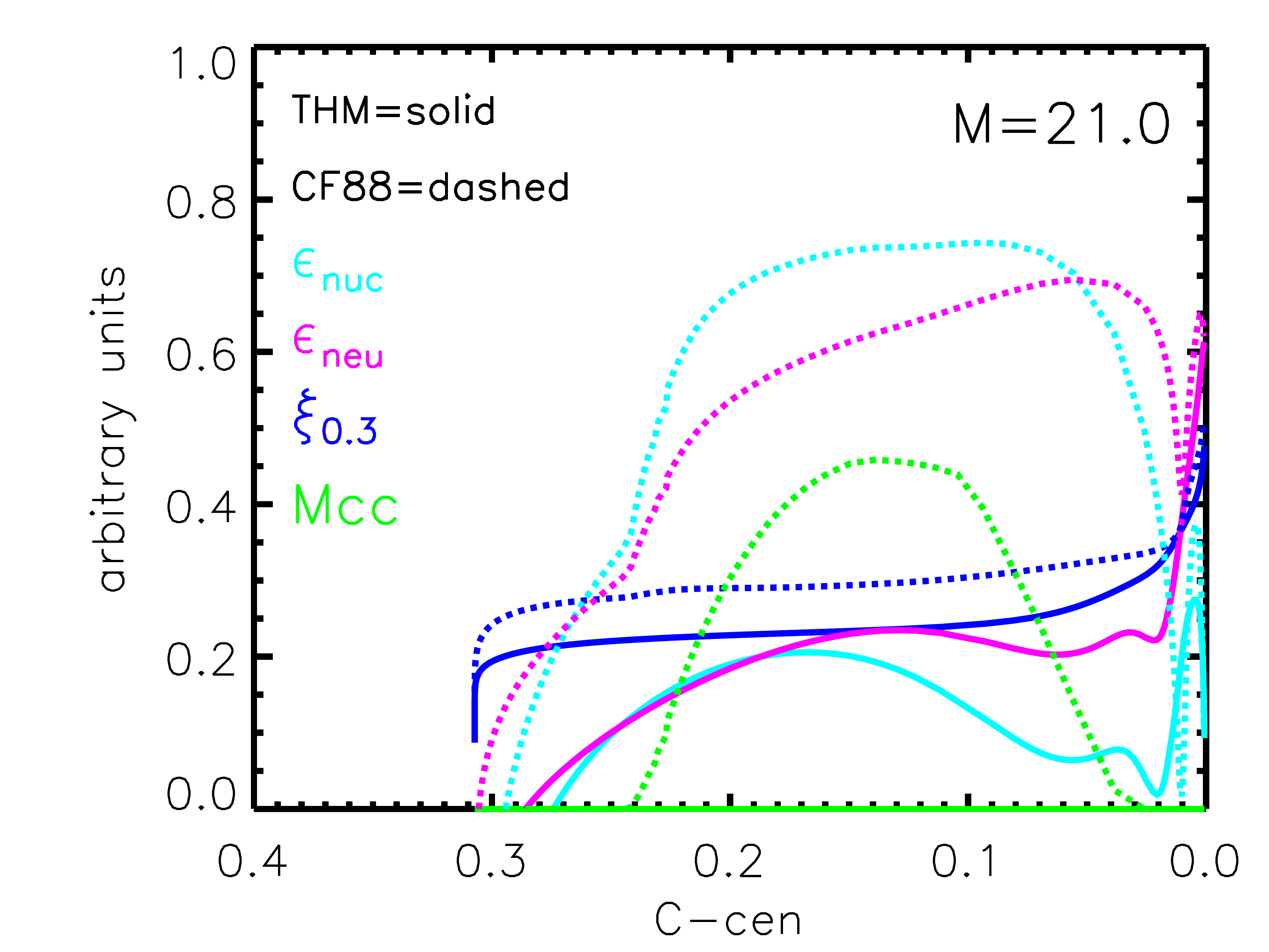}  {0.45\textwidth}{(b)}}
\caption{Run of of the central nuclear energy rate (cyan), the central neutrino energy losses (magenta), the maximum size of the convective core (green), the $\xi_{0.3}$ (blue) parameter, all as a function of the central C abundance. The solid and dashed lines refer, respectively, to the THM and CF88 models. Panels a) and b) refer, respectively, to the 15\msun~ and the 21\msun.\label{fig:carburan}}
\end{figure}

\begin{figure}[ht!]
\epsscale{1.1}
\plotone{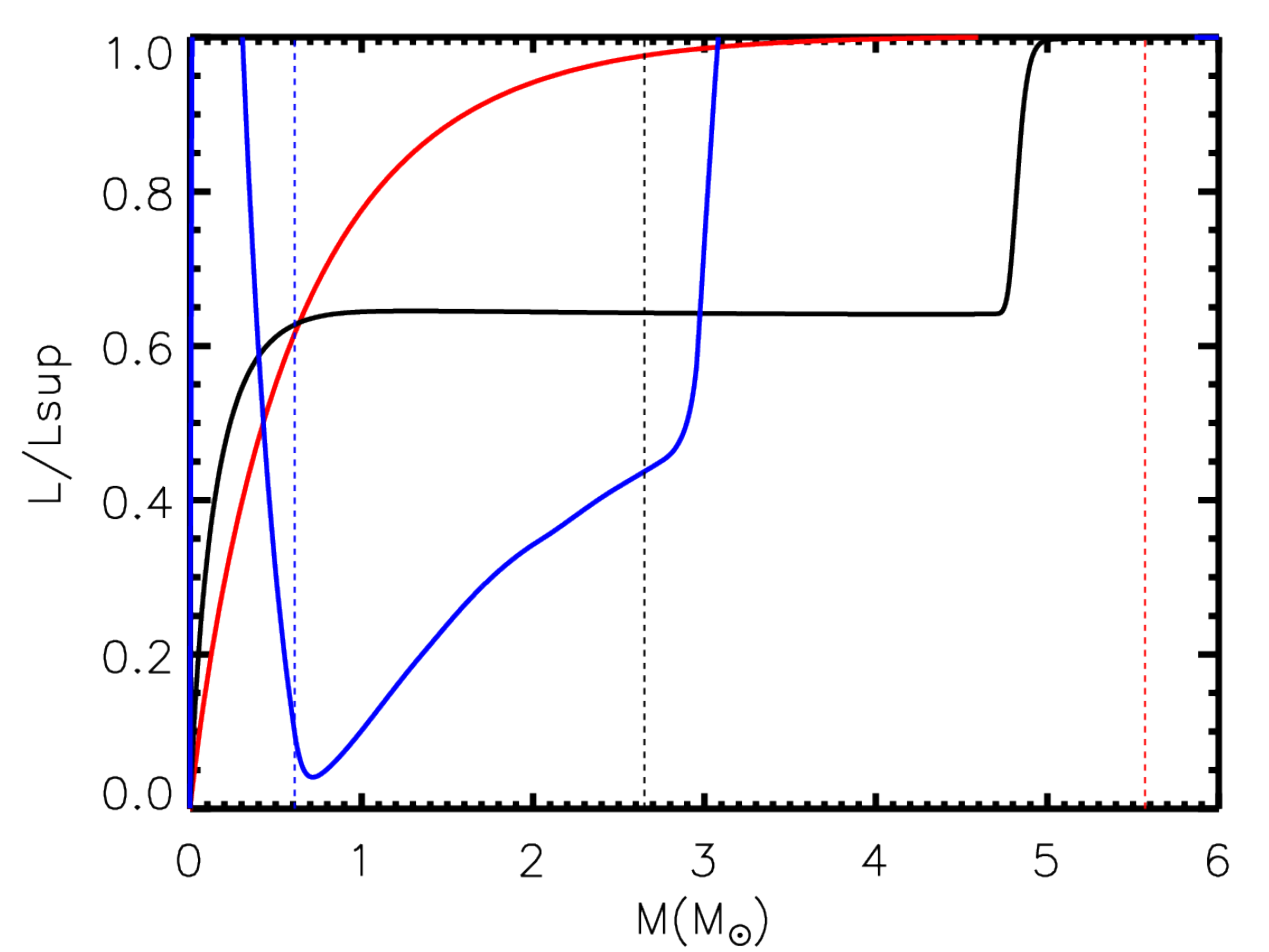}
\caption{Typical luminosity profiles of a massive star in H (red), He (black) and C burning (blue). The vertical dashed lines mark, in the three cases, the location of the border of the convective core in the three cases. \label{fig:lumi}}
\end{figure}

The neutrino energy losses $\rm(\epsilon_\nu)$ (mainly coming from $\rm e^{+}~e^{-}$ annihilation) that activate at temperatures larger than $\sim800$ MK, show a strong dependence on temperature and density. In Figure \ref{fig:conftcroc} black lines of constant $\rm log_{10}(\left|\epsilon_\nu\right|)$ are shown, together with the path followed by the 15\msun~(solid lines) and the 21\msun~(dashed lines) for both the CF88 (red lines) and the THM (blue lines) cases. The filled dots mark the beginning of the carbon burning, that we have defined as the model in which the central C abundance drops by 1\% with respect to the amount left by the He burning. Both masses show that an increase of the \nuk{C}{12}+\nuk{C}{12} nuclear reaction rate leads to a shift of the central C ignition towards lower temperatures and densities, where the neutrino energy losses are significantly lower. This has a profound influence on the evolution of the stars. The first reason is that the whole inner core of the stars is less compact at the beginning of the core C burning; this is well visible in Figure \ref{fig:carburan} where the run of the compactness parameter $\xi$ (blue lines), evaluated at 0.3\msun, is shown for both the 15 and the 21\msun, as a function of the central C abundance. It is evident that the THM models (solid lines) remain more expanded throughout the whole central C burning with respect to the CF88 ones (dashed lines). The second reason is that the lower the energy losses, the lower the amount of nuclear energy the C burning must supply to the star to keep it in equilibrium. This obviously means that the same amount of fuel lasts longer and hence it trivially explains why the adoption of the new rate increases the lifetime of the stars in this mass interval. 

A little bit less trivial is the reason why the run of the maximum size of the convective core with the initial mass scales as shown in Figure \ref{fig:cbur}. The two panels in Figure \ref{fig:carburan} show some key quantities for the 15 and the 21\msun. The 15\msun~ is representative of the range of masses in which the THM models show a convective core larger than in the CF88 case while the 21\msun~ is representative of the range of models in which the adoption of the new rate inhibits the formation of a convective core. In both panels the solid and dashed lines refer, respectively, to the THM and the CF88 models. In the same Figure are shown the run of the central nuclear energy generation rate (cyan lines), the central neutrino energy losses rate (magenta lines), the mass size of the convective core (green lines) and the compactness parameter $\xi$ (blue lines), evaluated at 0.3\msun, as a function of the central C abundance. In the 15\msun~ the nuclear energy production starts before the neutrino energy losses in both the CF88 and THM models; in fact the cyan line increases well before the magenta one for both the CF88 (dashed) and THM (solid) evolutionary sequences. This allows in both cases the formation of a convective core as soon as the C burning starts. However this does not explain directly why the convective core is larger when the higher rate is adopted. To understand this, it must be reminded that the presence of the neutrinos changes drastically the physics that controls, at each time, the size of the convective core. Both in H and He burning the luminosity increases moving from the center outward and then remains roughly flat once the nuclear energy generation drops to zero. This means that the balance between radiative and adiabatic gradient (that controls the border of the convective core if the Schwarzchild criterion is adopted) is the result of the smooth interplay among all the physical quantities that enter the game ($\rm \nabla_{rad}\propto \kappa P L / (M T^4)$ and $\rm \nabla_{ad})$. In C burning the situation is completely different because the neutrino energy losses are efficient on a much wider mass interval than the nuclear energy production (that is strongly concentrated in the very center). The result is that the luminosity profile raises steeply very close to the center of the star and then drops down dramatically as the neutrino energy losses erode the energy flux coming from the center. It is this drastic luminosity drop, determined by the neutrino losses, that controls the border of the convective core in these conditions. Since the adoption of the higher rate forces the C burning to occur in an environment in which the neutrino energy losses are less efficient than in the standard (CF88) case, the radiative gradient remains above the adiabatic one up to larger masses. Figure \ref{fig:lumi} shows a sketch of the typical luminosity profile a massive star has in H, He and C burning.
If we now turn to the 21\msun~ the scenario is quite different. In this case, the neutrino energy losses dominate the C burning since the beginning, preventing the early growth of convective instabilities, panel b) in Figure \ref{fig:carburan}. Only the continuous increase of the central temperature allows the nuclear burning to significantly overcome the neutrino energy losses (cyan dotted versus magenta dotted in the Figure). Hence in this case the convective core forms only after a while, when the central C abundance drops to 0.24 or so. Also in the THM case the neutrino energy losses activate before the nuclear burning (solid cyan and magenta lines in the Figure) but in this case, since the evolution proceeds in an environment where the neutrino losses are less efficient, the nuclear energy never increases enough to form a convective core. This is evident by comparing the nuclear energy rates produced by the CF88 and the THM models (solid versus dotted cyan in panel b) of Figure \ref{fig:carburan}): the adoption of the higher rate implies a much lower request of nuclear energy, and the consequence is that a convective core never develops. Above the 24\msun~ or so, both sets of models do not develop a convective core because the nuclear energy gain does not overcome significantly any more the neutrino energy losses (in C burning, of course).

\section{The shell carbon burning} \label{sec:carshe}
\begin{figure}[ht!]
\gridline{\fig{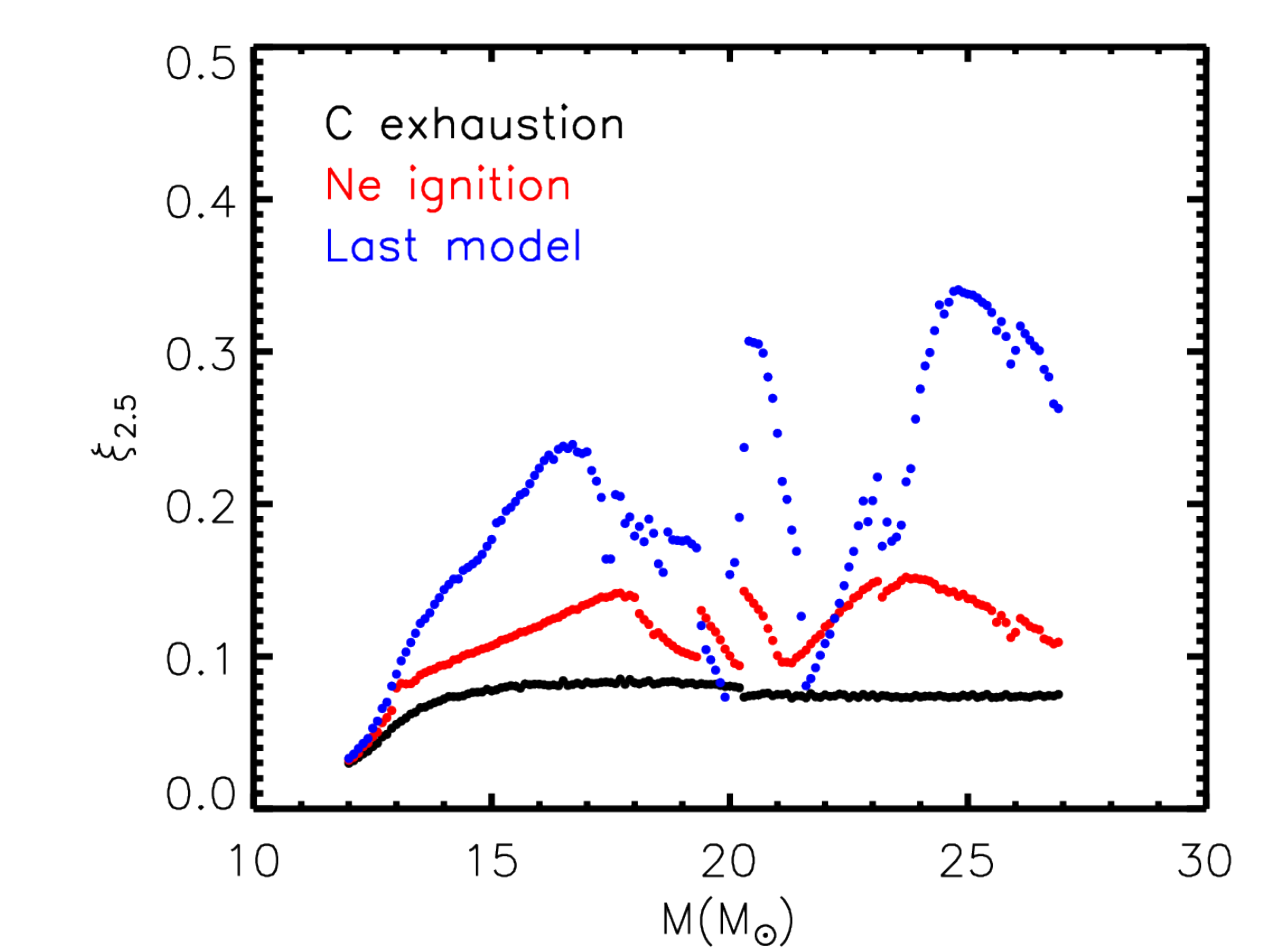}  {0.45\textwidth}{a) Compactness evaluated at the end of central C burning (black dots), at the beginning of the Ne burning (red dots) and at the onset of the core collapse (blue dots)}}
\gridline{\fig{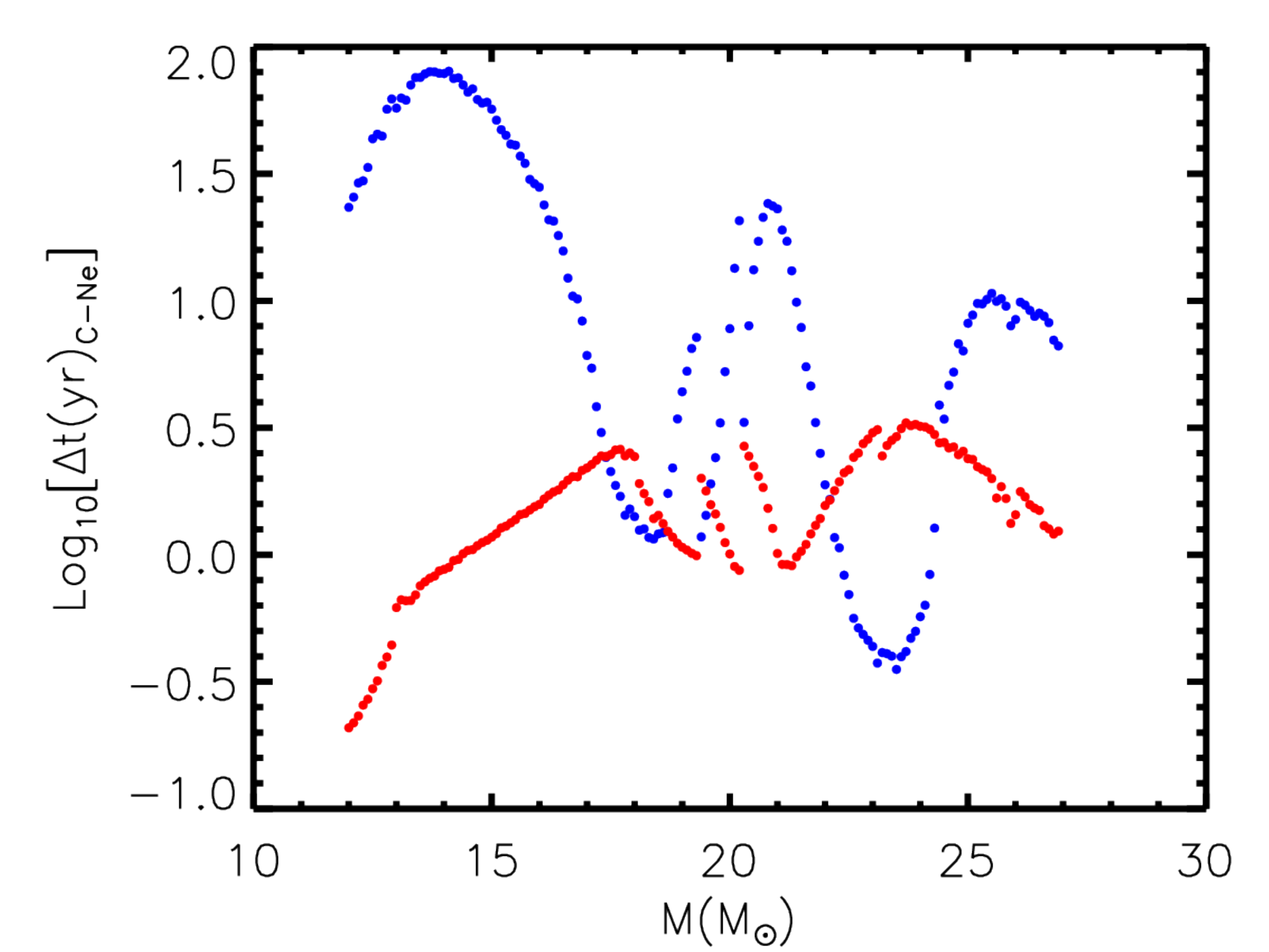}  {0.45\textwidth}{b)timescale of the evolution from the end of the central C burning to the beginning of the Ne burning (blue dots) and the $\xi_{2.5}$ evaluated at the beginning of the Ne burning (red dots) in an arbitrary scale}}
\caption{\label{fig:vacsi}}
\end{figure}

Once C is heavily depleted in the center the core contracts, heats and the burning moves outward, where matter is still C rich. At this stage the compactness $\xi_{2.5}$ does not present any specific feature: it shows almost no dependence on the initial mass (black dots in Figure \ref{fig:vacsi}, panel a). All the key features present at the beginning of the core collapse (blue dots in Figure \ref{fig:vacsi}, panel a)  show up very clearly already at the beginning of Ne burning, though obviously the other advanced burning (Ne, O and Si) contribute to the final shape of the compactness. This is readily visible in Figure \ref{fig:vacsi} where the red dots in panel a) show the scaling of compactness with the mass at the beginning of Ne burning. The main features that show up at this stage are qualitatively maintained up to the onset of the core collapse. In principle, since the mass of the CO core increases and the C mass fraction left by the He burning decreases with the initial mass, one would expect the final compactness to show a steady featureless increase with the initial mass. But this is not the case. Panel b) in Figure \ref{fig:vacsi} shows the time interval that elapses between the end of the central C burning and the Ne ignition, $\rm \Delta t$(C-Ne) - blue dots, as a function of the initial mass. For convenience, in this panel it is shown again the compactness at the beginning of the Ne burning on an arbitrary vertical scale (red dots). It is evident that the time interval between the C exhaustion and the Ne ignition does not decrease continuously (as expected on the basis of the above arguments) but presents features that are highly correlated with the compactness. In particular, the slow down of the contraction of the CO correlates quite well with a decrease of the compactness.

Moving from the 12\msun~up to the 17.3\msun, all stars present a C convective core followed by a very extended C convective shell that vanishes somewhat before the Ne ignition. A second convective shell episode occurs much later, after the central O burning. Figure \ref{fig:kipp1} shows in panel a) the Kippenhahn diagram of a 15\msun~ as a typical example of the stars in this mass interval. The compactness increases monotonically in all this mass interval because the convective episodes are basically the same (qualitatively) in all stars, so that they do not alter the expected direct scaling of the compactness with the CO core mass. It must be however noted that the time interval $\rm \Delta t$(C-Ne) in this mass range reaches a maximum (around 14.5\msun) and then decreases; since this quantity scales directly with the efficiency of the C burning, this is an indication of a non monotonic behavior of the power of both the C convective core and the C convective shell as a function of the initial mass in this mass interval.

As the initial mass increases, the second C convective shell progressively anticipates its formation and around 17.4\msun~ it becomes effective together to, or even before, the formation of the Ne convective core. The growth of this additional strong energy source before Ne ignition slows down the contraction of the star between the end of the C burning and the Ne ignition and the compactness reduces accordingly. Also the $\rm \Delta t$(C-Ne) increases, at least in the mass interval 18.3-19.4\msun. Of course one should expect only a qualitative correlation between compactness and timescale of contraction because of the complex and highly non linear interplay among the many components of the star that contribute to its evolution.

The early formation of a second C convective shell (i.e. before the Ne ignition) vanishes in stars with initial mass greater than $\rm \sim 19.4~M_\odot$ because of the progressive reduction of the efficiency of the C burning (due again to the continuous increase of the CO core mass and the decrease of the C left by the He burning). As a consequence, the contraction of the CO core speeds up, which means a sudden decrease of the time scale of contraction, and hence a sharp increase of the compactness shows up. A second sharp discontinuity in the compactness profile occurs at 20.3\msun. This feature is easily understandable because it coincides with the mass above which the central C burning does not form any more a convective core. The lack of a convective core changes qualitatively the speed at which the contraction of the core occurs, and this leads to the sharp discontinuity visible at 20.3\msun. The counter effect of this sharper contraction leads to a progressive increase of the power of the C convective shell so that the compactness progressively reduces for a while. However, the main engine that drives the contraction slowly takes again the control of the compactness, and around 21\msun~ it starts raising again. The next (last) turn over in the compactness profile occurs at roughly 24\msun~when only one powerful C convective shell, able to slow down the contraction of the CO core, forms. The shape sculpted by the various (more or less efficient) C convective episodes changes quantitatively between the Ne burning and the onset of the collapse, in the sense that the minima and maxima already visible at the beginning of the Ne burning become much more pronounced but do not shift significantly in mass in the advanced burning. For example the maximum present at 17.8\msun in the shape of the compactness at the Ne ignition, shifts at 16.5\msun in the final configuration.
A comparison between the compactness of the models at the Ne ignition and at the onset of the collapse shows that there are two small mass ranges, one around 19.4-20\msun and the other one around 21.5-22.2\msun, in which the final compactness is lower than that at the beginning of the Ne burning. In other terms these stars are more expanded at the end of the hydrostatic evolution than at the Ne ignition. This "atypical" behavior is the consequence of the penetration of the C convective shell in the tail of the He profile (the outer border of the convective shell reaches the zone where the local He abundance is of the order of 0.03 by mass fraction). The fresh He engulfed in the C convective shell forces a burst of the C burning that, in turn, leads to the sudden expansion of the mantle (and hence a drop of the compactness) from which the star is not able to recover in the short time remaining before the collapse.

\section{Discussion and conclusions}

\begin{figure}[ht!]
\gridline{\fig{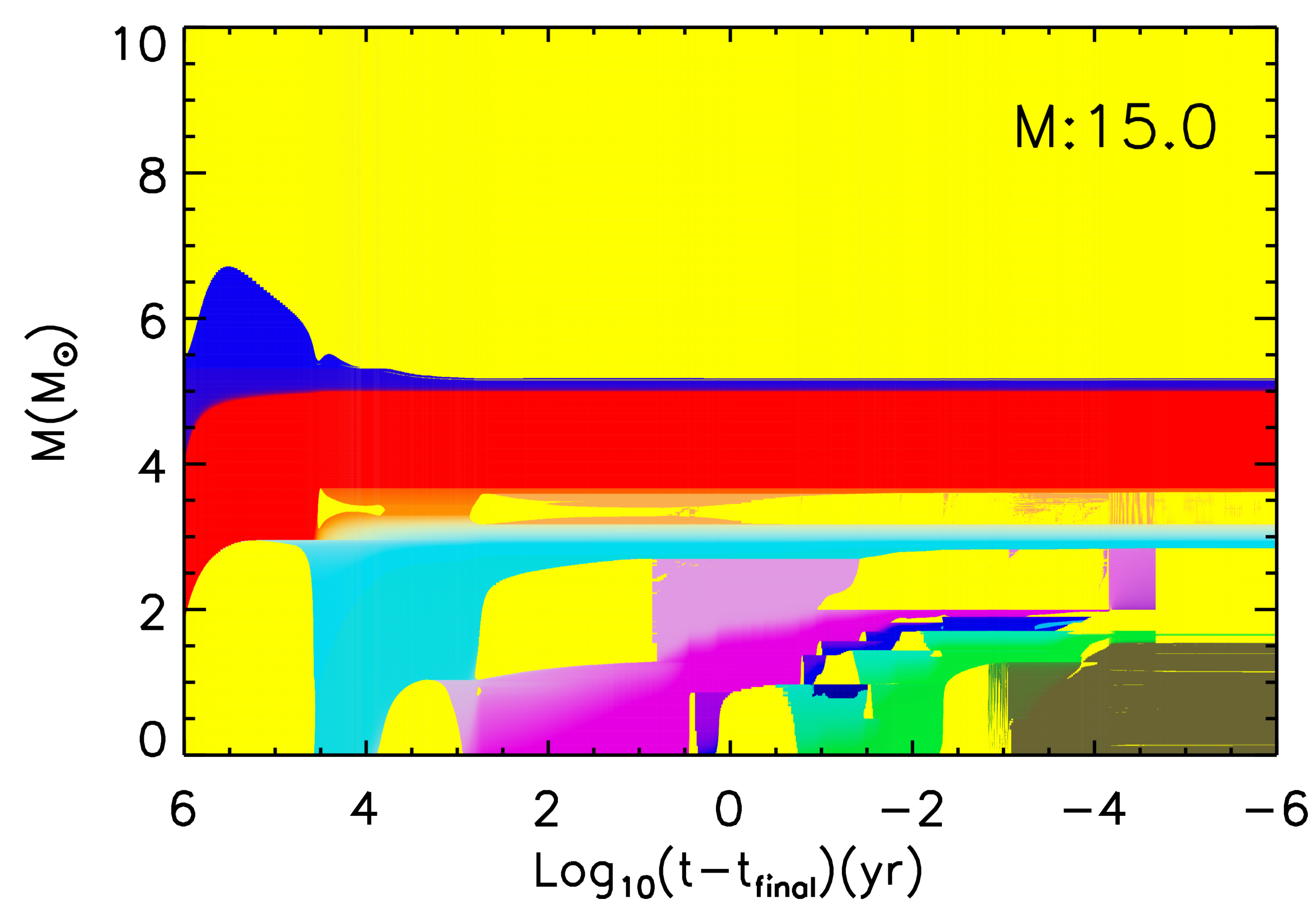}  {0.5\textwidth}{(a)}
          \fig{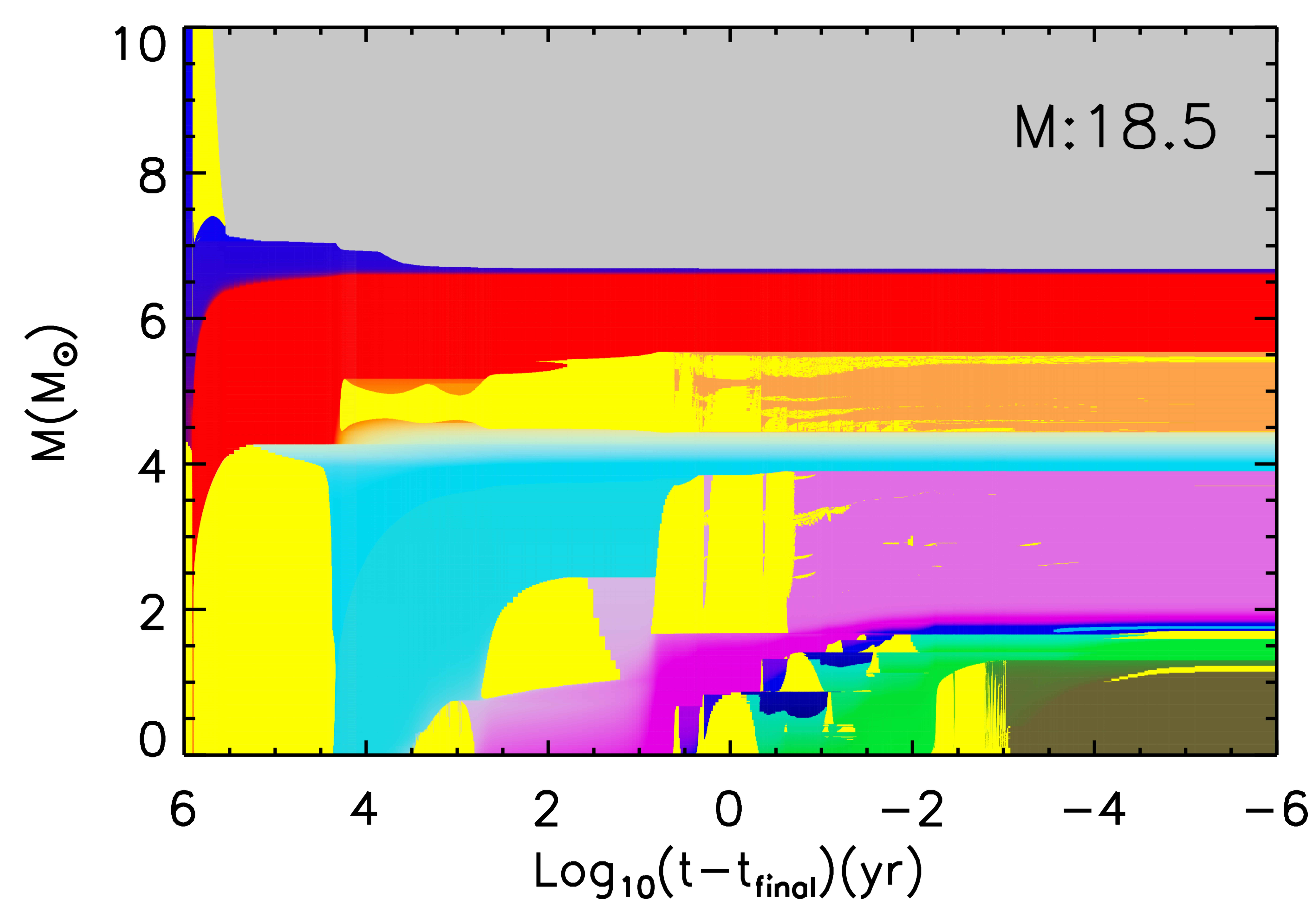}  {0.5\textwidth}{(b)}}
\gridline{\fig{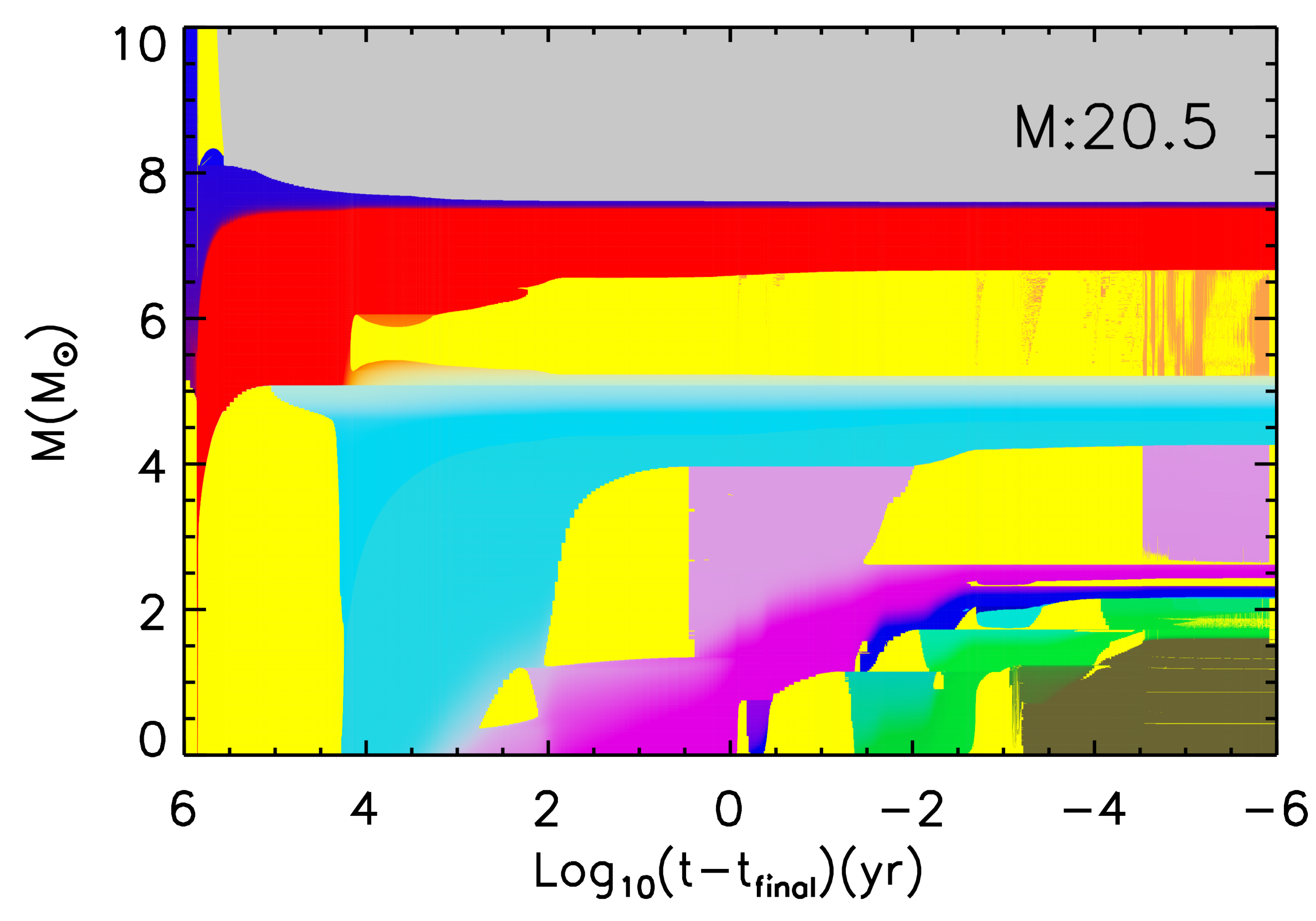}  {0.5\textwidth}{(c)}
          \fig{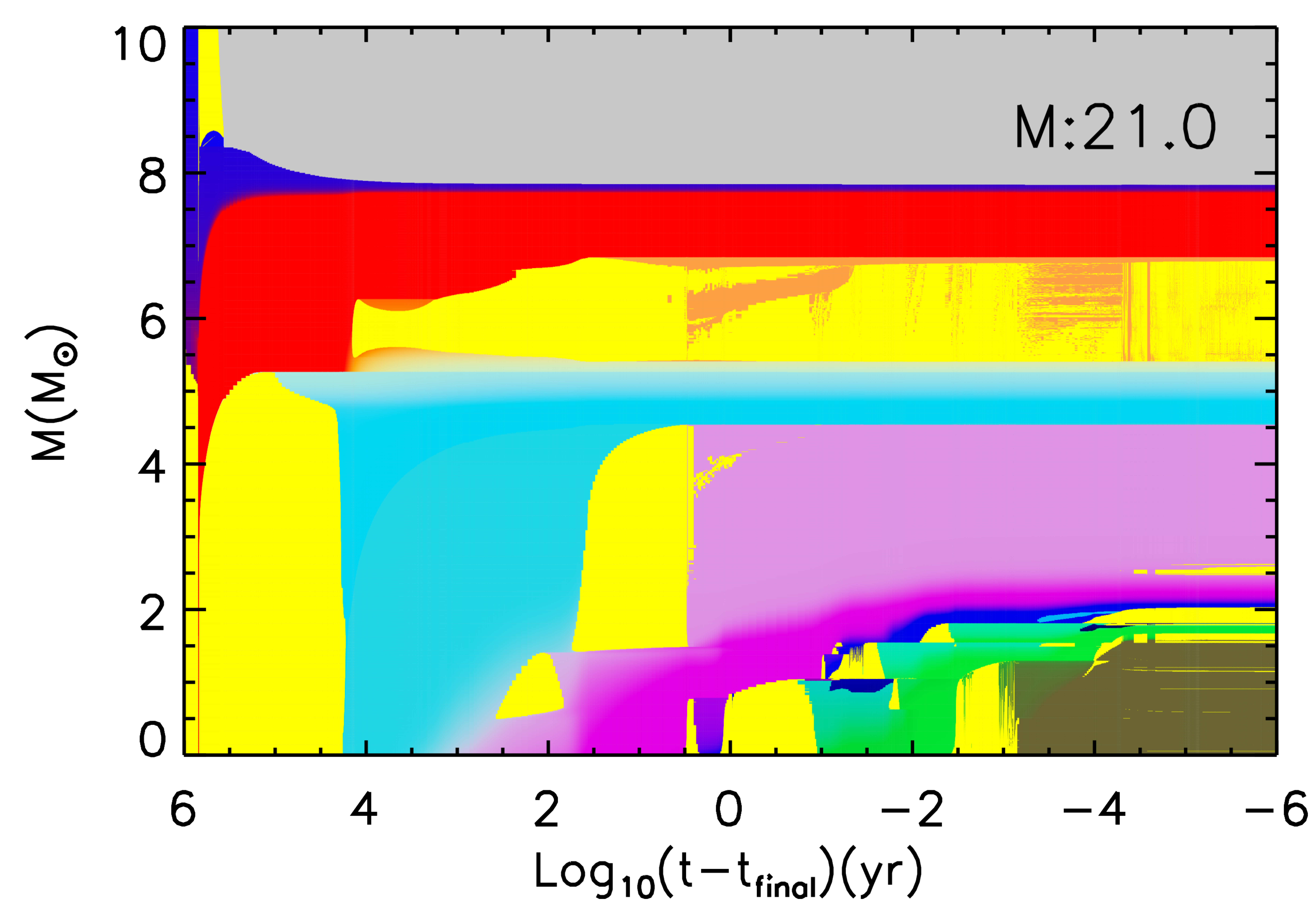}  {0.5\textwidth}{(d)}}
\gridline{\fig{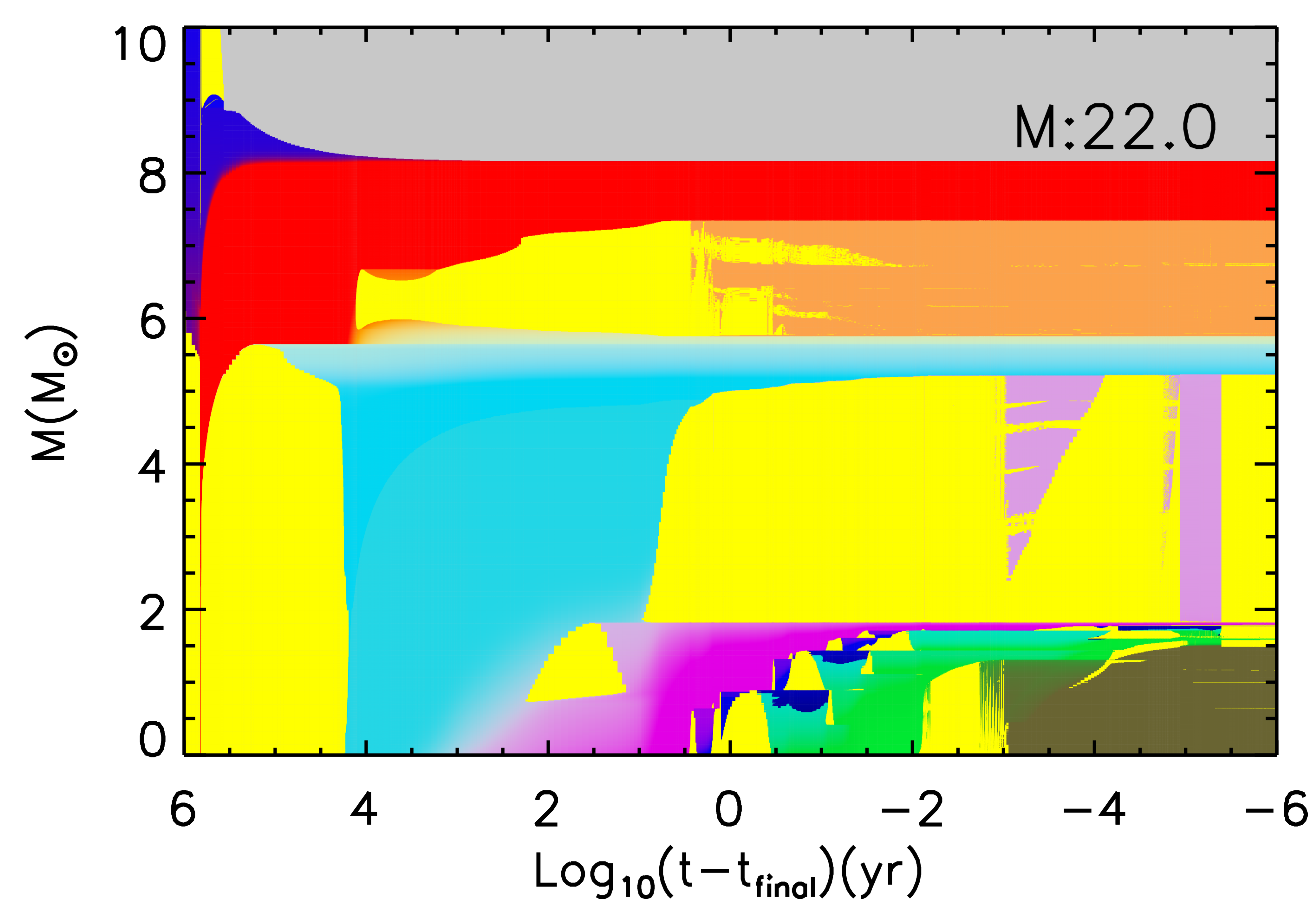}  {0.5\textwidth}{(e)}
          \fig{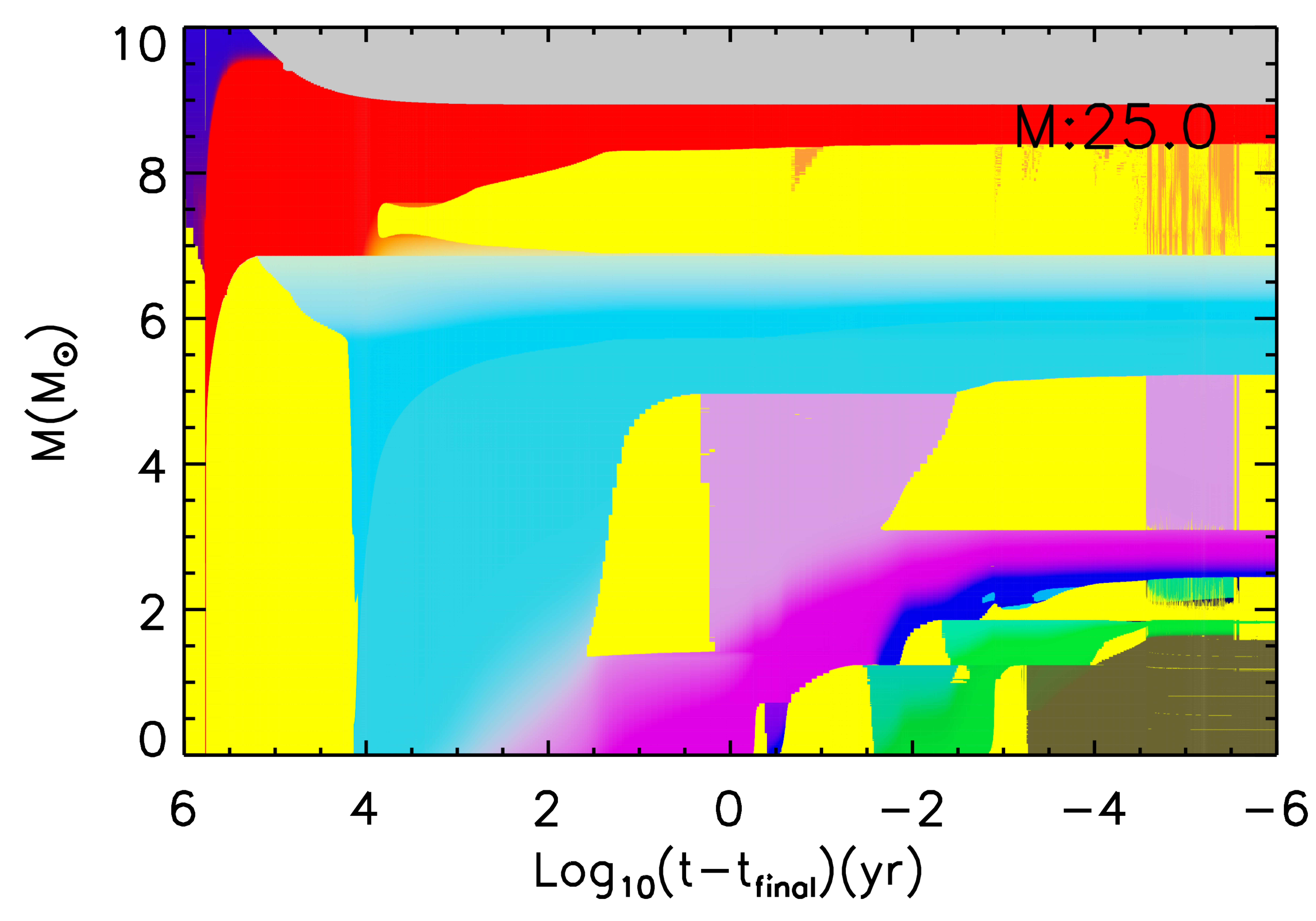}  {0.5\textwidth}{(f)}}
\caption{Kippenhanh diagrams of selected models. Models computed with the \nuk{C}{12}+\nuk{C}{12} nuclear cross section measured by \cite{thm18} with the Trojan Horse Method. The color coding is as follows: all the convective regions are in yellow, the H rich ones are blue and the He rich ones in red. The He exhausted zone is cyan if C is present, otherwise magenta if Ne is present or dark blue if only O is present. Si rich zones are green while the hashes of the Si burning are dark green. All the transition colors show intermediate chemical compositions.\label{fig:kipp1}}
\end{figure}

\begin{figure}[ht!]
\gridline{\fig{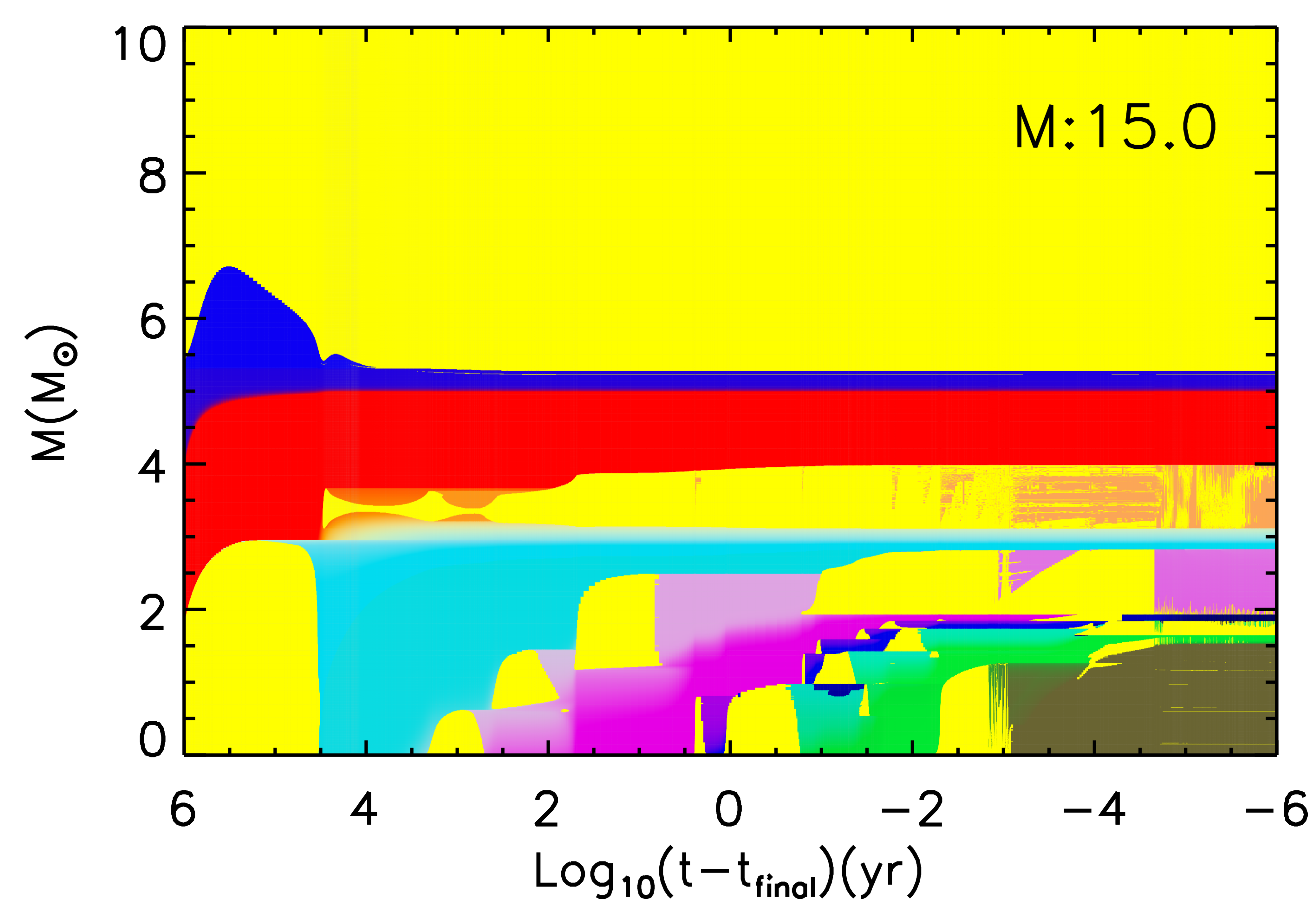}  {0.5\textwidth}{(a)}
          \fig{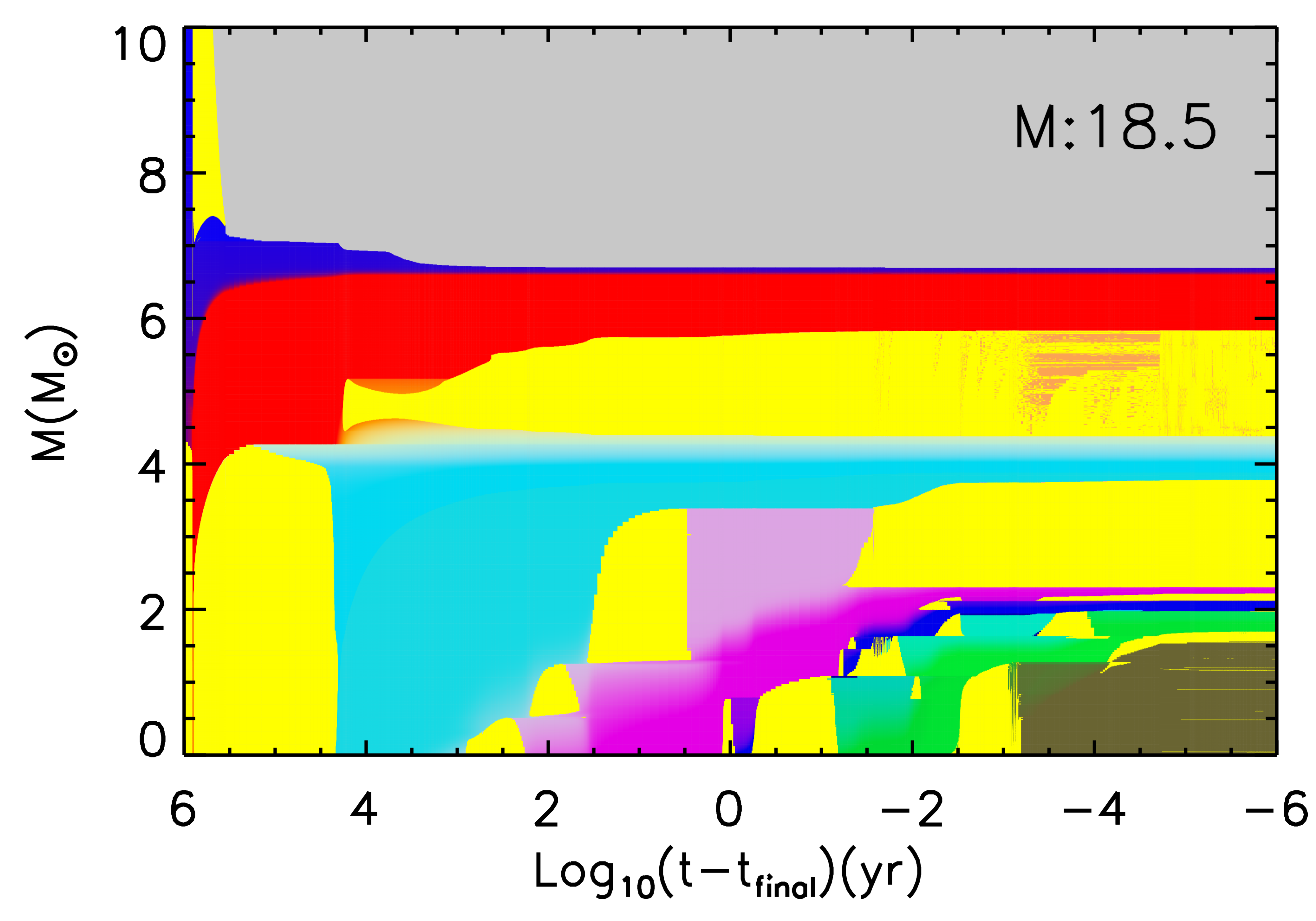}  {0.5\textwidth}{(b)}}
\gridline{\fig{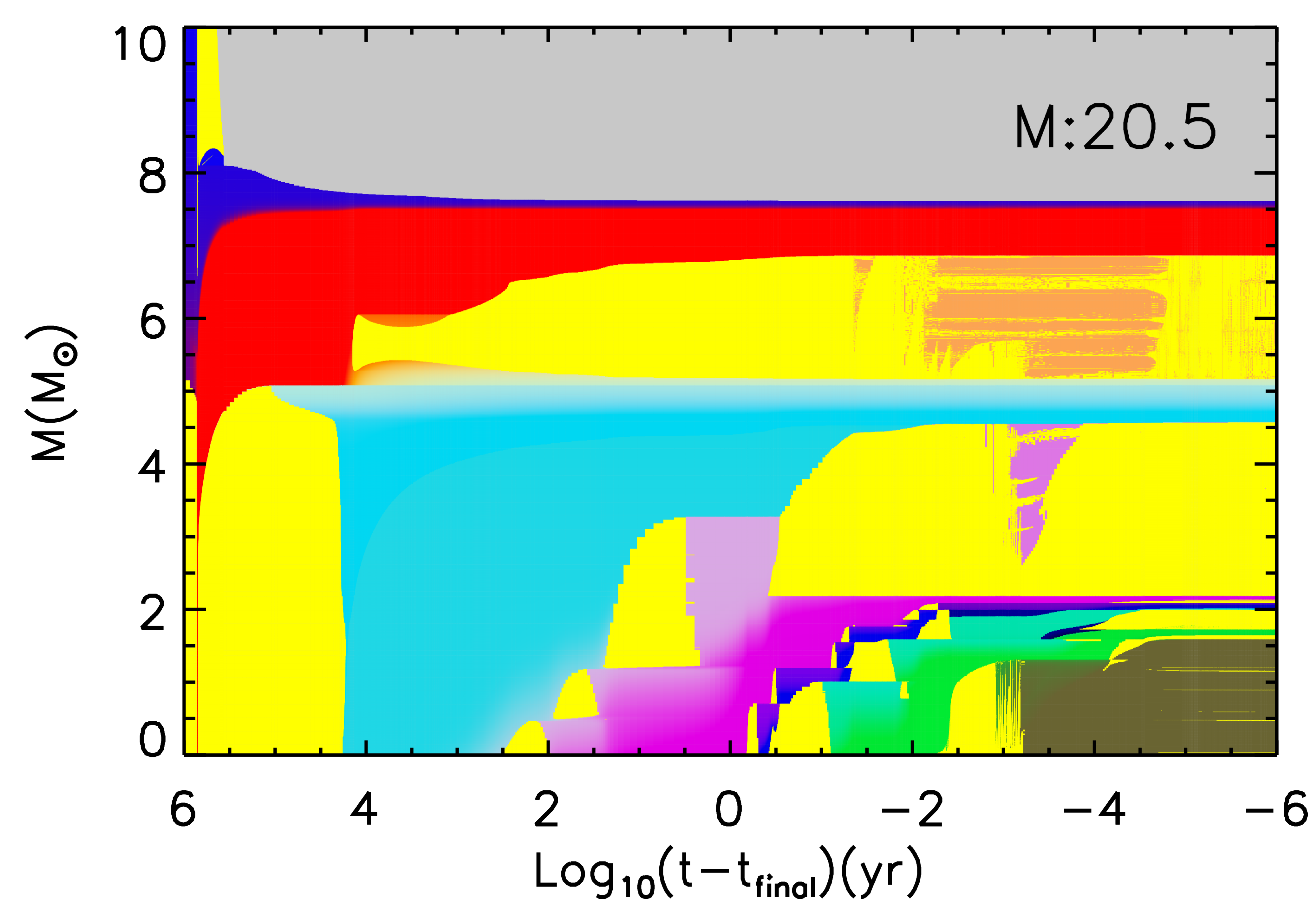}  {0.5\textwidth}{(c)}
          \fig{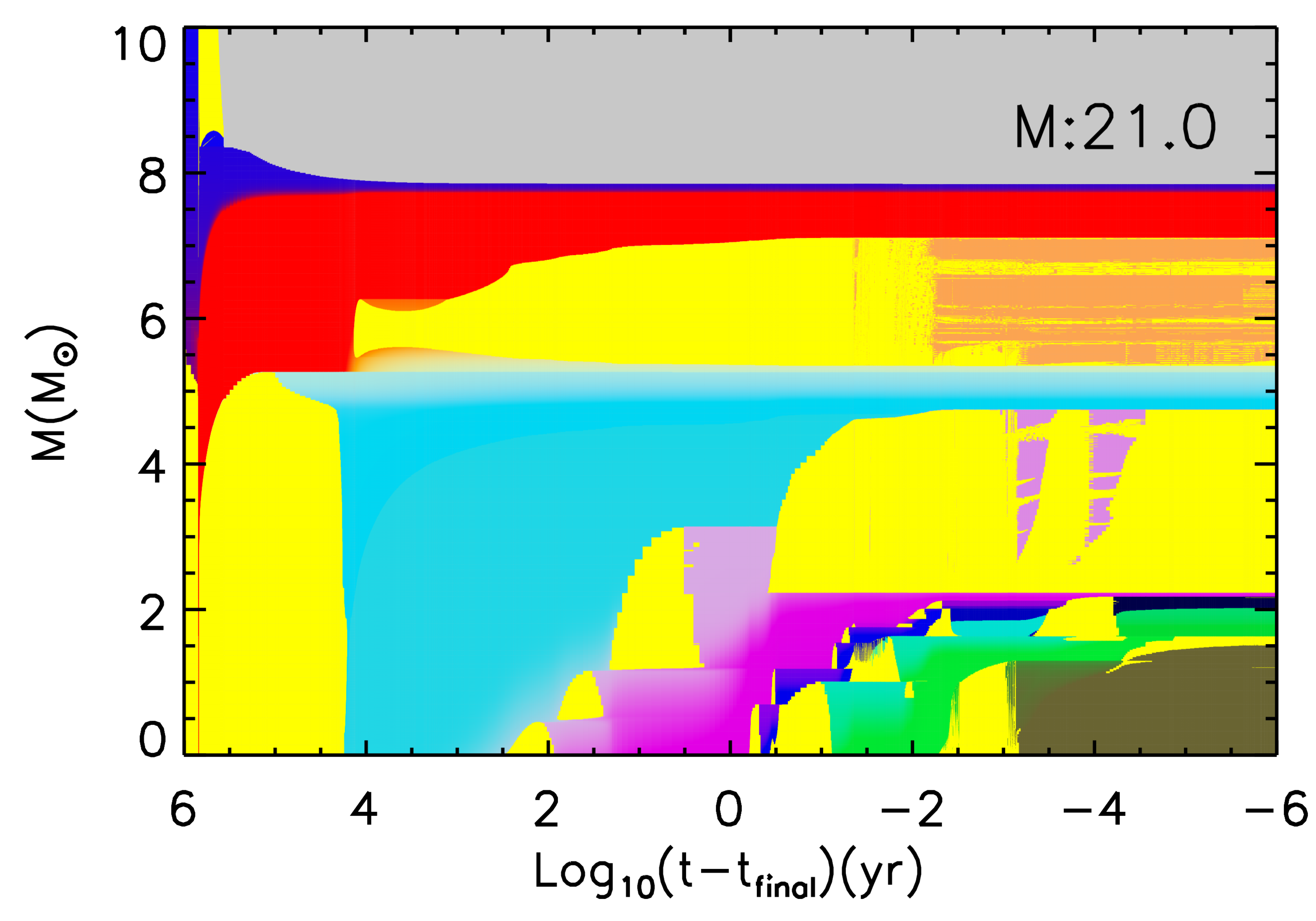}  {0.5\textwidth}{(d)}}
\gridline{\fig{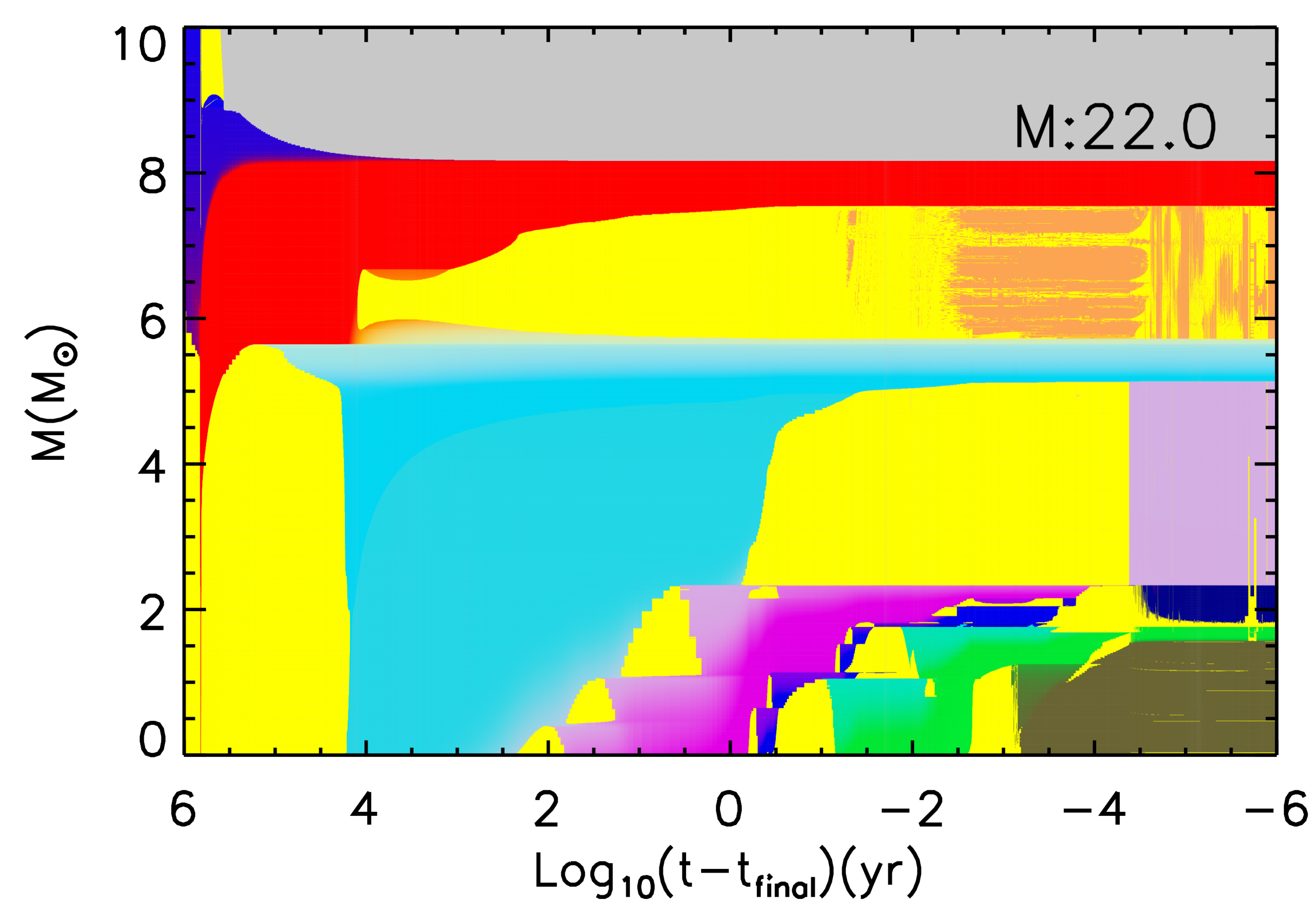}  {0.5\textwidth}{(e)}
          \fig{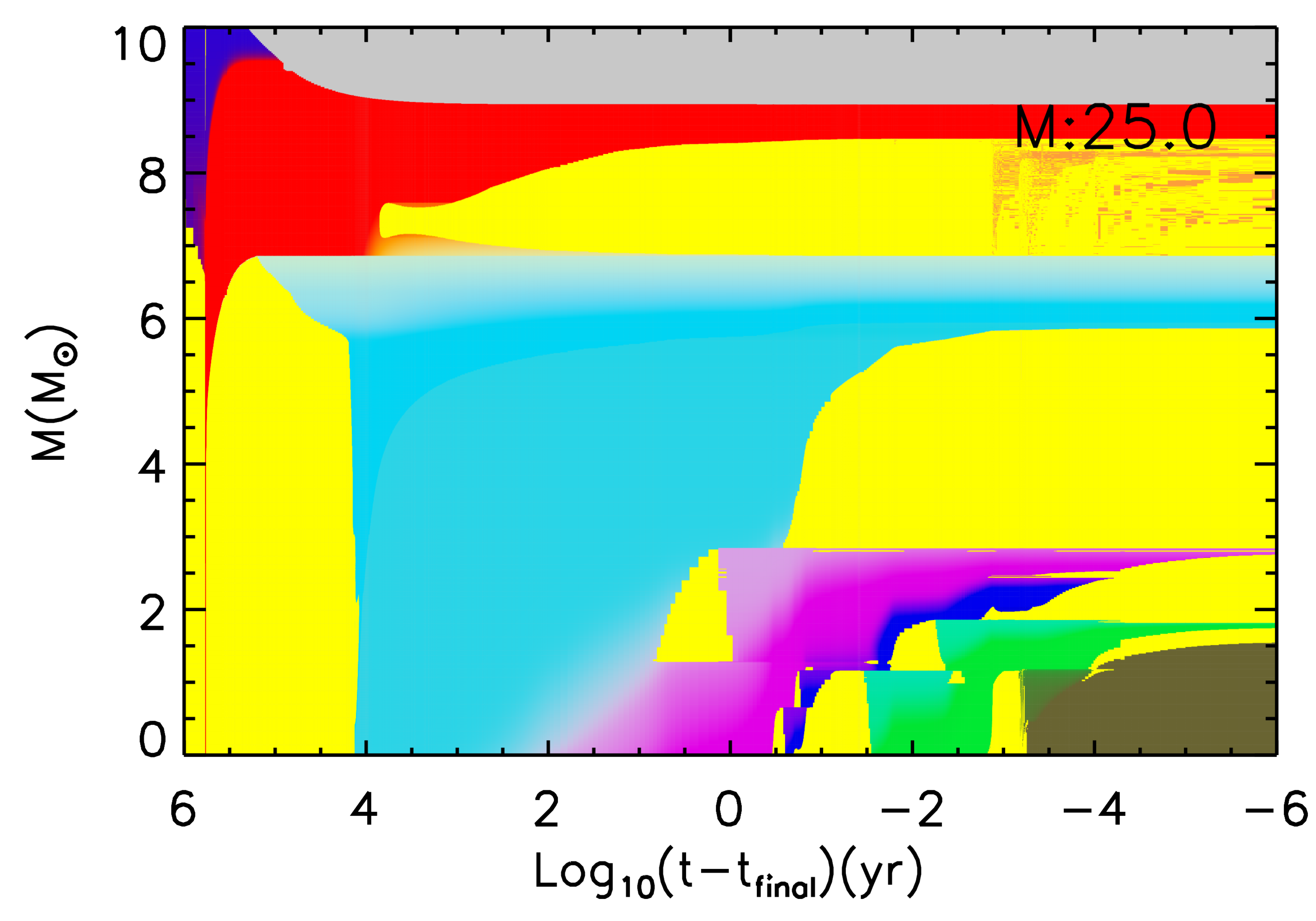}  {0.5\textwidth}{(f)}}
\caption{Kippenhanh diagrams of selected models. Models computed with the \nuk{C}{12}+\nuk{C}{12} nuclear cross section given by CF88. The color coding is as follows: all the convective regions are in yellow, the H rich ones are blue and the He rich ones in red. The He exhausted zone is cyan if C is present, otherwise magenta if Ne is present or dark blue if only O is present. Si rich zones are green while the hashes of the Si burning are dark green. All the transition colors show intermediate chemical compositions.\label{fig:kipp2}}
\end{figure}

In sections \ref{sec:carcen} and  \ref{sec:carshe} we discussed in detail the changes induced by the adoption of the new nuclear cross section on the C burning, both central and in shell. Analogous results were obtained by \cite{be12} and \cite{pi13}, who already found that the lifetime of a star in central C burning scales directly with the \nuk{C}{12}+\nuk{C}{12} nuclear cross section. They also found that an increase of this cross section reduces the number of convective episodes though each of them is on average more extended in mass. This result, already shown for example in Figure 2 of \cite{pi13}, is evident by comparing Figure \ref{fig:kipp1}, obtained by assuming the THM nuclear cross section, to Figure \ref{fig:kipp2} that shows the same masses but computed with the CF88 nuclear cross section: the adoption of a higher \nuk{C}{12}+\nuk{C}{12} nuclear cross section leads to a lower number but more extended C convective shell episodes. Since we needed a large number of models to have the right resolution in mass for the purposes of the present project, we were forced to adopt a small nuclear network that does not allow us to discuss the changes induced by the new nuclear cross section on the yields. However this kind of analysis may be found in \cite{ga07}, \cite{be12} and \cite{pi13}.

\begin{figure}[ht!]
\epsscale{1.1}
\plotone{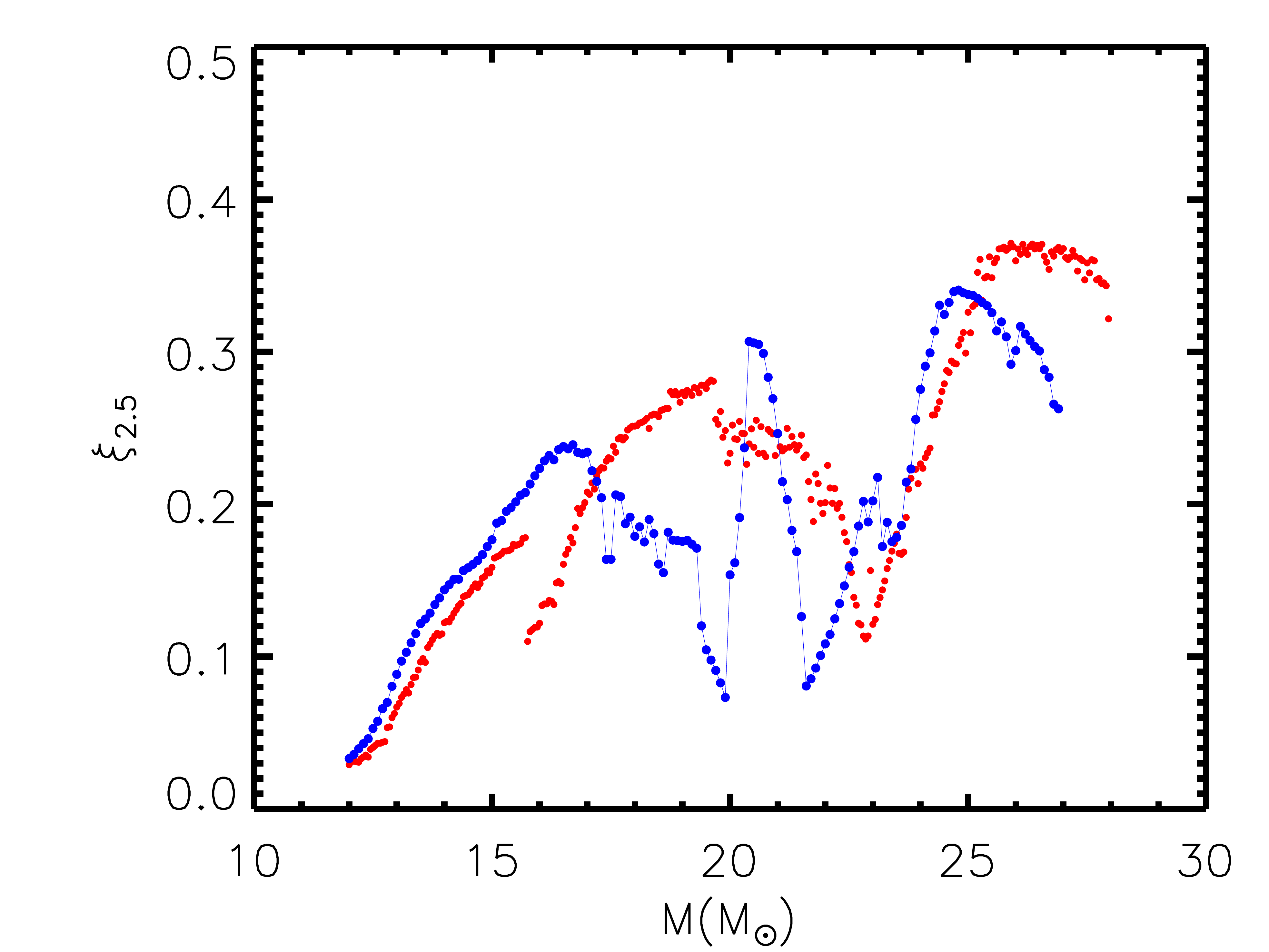}
\caption{Comparison between the final compactness obtained by adopting the \nuk{C}{12}+\nuk{C}{12} nuclear cross section provided by CF88 (red dots) and the new one measured by \cite{thm18} (blue dots).\label{fig:csi}}
\end{figure}

The differences in the efficiency of the C burning (induced by the adoption of different nuclear cross sections) lead also to a different scaling of the final compactness with the mass. Figure \ref{fig:csi} compares the trend we obtained in CL20 by adopting the CF88 \nuk{C}{12}+\nuk{C}{12} nuclear cross section (red dots) to that obtained in this paper where we adopted the recent new measurement of the \nuk{C}{12}+\nuk{C}{12} nuclear cross section published by \cite{thm18}. The comparison between the results obtained with the two different nuclear cross sections shows that there are significant differences between the two sets of models, the new ones showing a remarkable lower $\xi_{2.5}$ with respect to older ones in some mass intervals, like between 17-20\msun, 21-23\msun and above 25\msun. Though we cannot quantify the relation between explodability and compactness for the reasons discussed in the Introduction, it seems that in some mass intervals the explosion could proceed quite differently in the two cases. It would be interesting to follow with a 3D code two models having the same mass, e.g. around 20\msun, where the differences between the two cases are quite extreme. Let us eventually remark that also the new trend shows features but no significant scatter around the mean path.

The present results may be summarized by saying that the three peaks present in the final scaling of the compactness with the initial mass are due to: the presence of a convective core followed by one (the raise up to the first maximum) or two convective shells (the drop after the first maximum); a sharp narrow peak that extends over roughly 2\msun~ corresponds to the readjustment of the structure due to the disappearance of the convective core. At the beginning, the lack of a convective core forces the CO core to contract more steadily but such a strong compression activates, after a while, a strong convective shell that operates in the opposite direction, slowing down the contraction. After this readjustment of the structure due to the transition from a convective to a radiative core carbon burning, the mass size of the CO core plus the amount of C left by the He burning take again the control of the contraction which therefore scales again directly with the initial mass. The formation of a single very powerful C convective shell, above $\sim26$\msun, marginally reduces the final compactness of a star.
Unfortunately, these models do not allow an analysis of the differences in the nucleosynthesis produced by the adoption of this new nuclear cross section because, given the very large number of models, we were forced to adopt a small nuclear network (see CL20). We will address the influence of this new nuclear cross section on the final yields in a forthcoming paper. 

The evolutionary tracks are available upon request and will be added very shortly to our online repository (http://orfeo.iaps.inaf.it/).


\end{document}